\documentclass[acmsmall]{acmart}
\usepackage{enumerate}
\usepackage{multirow}
\usepackage{bbding}
\usepackage{makecell}
\usepackage{xcolor}
\colorlet{RED}{red}

\usepackage{graphicx}
\usepackage{subcaption}

\usepackage{tcolorbox}

\AtBeginDocument{%
  }

\acmJournal{JACM}
\begin{document}

\title{Survey of Code Search Based on Deep Learning}


\author{Yutao Xie}
\affiliation{%
  \institution{Peking University \& International Digital Economy Academy}
  \streetaddress{No. 5 Shihua Road}
  \city{Futian District}
  \state{Shenzhen}
  \country{China}
  \postcode{518017}
}
\email{yutaoxie@idea.edu.cn}

\author{Jiayi Lin}
\authornotemark[1]
\affiliation{%
  \institution{International Digital Economy Academy}
  \streetaddress{No. 5 Shihua Road}
  \city{Futian District}
  \state{Shenzhen}
  \country{China}
  \postcode{518017}
}
\email{jiayilin1024@gmail.com}

\author{Hande Dong}
\affiliation{%
  \institution{International Digital Economy Academy}
  \streetaddress{No. 5 Shihua Road}
  \city{Futian District}
  \state{Shenzhen}
  \country{China}
  \postcode{518017}
}
\email{donghd66@gmail.com}

\author{Lei Zhang}
\affiliation{%
  \institution{International Digital Economy Academy}
  \streetaddress{No. 5 Shihua Road}
  \city{Futian District}
  \state{Shenzhen}
  \country{China}
  \postcode{518017}
}
\email{leizhang@idea.edu.cn}

\author{Zhonghai Wu}
\affiliation{%
  \institution{Key Lab of High Confidence Software Technologies (MOE), Peking University}
  \streetaddress{No. 5 Yiheyuan Road}
  \city{Haidian District}
  \state{Beijing}
  \country{China}
  \postcode{100871}
}
\email{wuzh@pku.edu.cn}

\renewcommand{\shortauthors}{Xie et al.}

\begin{abstract}
  Code writing is repetitive and predictable, inspiring us to develop various code intelligence techniques. 
  This survey focuses on code search, that is, to retrieve code that matches a given natural language query by 
  effectively capturing the semantic similarity between the query and code.  
  Deep learning, being able to extract complex semantics information, has achieved great success in this field. 
  Recently, various deep learning methods, such as graph neural networks and pretraining models, 
  have been applied to code search with significant progress. 
  Deep learning is now the leading paradigm for code search. 
  In this survey, 
  we provide a comprehensive overview of deep learning-based code search. 
  We review the existing deep learning-based code search framework
  which maps query/code to vectors and 
  measures their similarity. 
  Furthermore, we propose a new taxonomy to 
  illustrate the state-of-the-art deep learning-based code search 
  in a three-steps process:
  query semantics modeling, code semantics modeling, and 
  matching modeling which involves the deep learning model training.
  Finally, we suggest potential avenues for future research in this promising field.
\end{abstract}

\begin{CCSXML}
<ccs2012>
   <concept>
       <concept_id>10002944.10011122.10002945</concept_id>
       <concept_desc>General and reference~Surveys and overviews</concept_desc>
       <concept_significance>500</concept_significance>
       </concept>
   <concept>
       <concept_id>10011007</concept_id>
       <concept_desc>Software and its engineering</concept_desc>
       <concept_significance>500</concept_significance>
       </concept>
   <concept>
       <concept_id>10011007.10011006</concept_id>
       <concept_desc>Software and its engineering~Software notations and tools</concept_desc>
       <concept_significance>300</concept_significance>
       </concept>
   <concept>
       <concept_id>10011007.10011074</concept_id>
       <concept_desc>Software and its engineering~Software creation and management</concept_desc>
       <concept_significance>100</concept_significance>
       </concept>
 </ccs2012>
\end{CCSXML}

\ccsdesc[500]{General and reference~Surveys and overviews}
\ccsdesc[500]{Software and its engineering}
\ccsdesc[300]{Software and its engineering~Software notations and tools}
\ccsdesc[100]{Software and its engineering~Software creation and management}

\keywords{code search, code understanding, natural language processing, deep learning, pre-training}

\setcopyright{acmlicensed}
\acmJournal{TOSEM}
\acmYear{2023} \acmVolume{1} \acmNumber{1} \acmArticle{1} \acmMonth{1} \acmPrice{15.00}\acmDOI{10.1145/3628161}

\maketitle

\section{Introduction}
Programming is the process of using a programming language to write code that performs a specific function. 
Despite being a creative endeavor, 
programming also exhibits repetitive and predictable attributes \cite{hindle2016naturalness}, with many codes already implemented to some extent. 
By analyzing historical codes, it is possible to anticipate the content of the code to be written. 
Building on this insight, 
code intelligence techniques such as code search, code completion, and code generation can be employed to enhance the productivity of software developers. 
In recent years, 
the application of deep learning in the field of code intelligence has led to remarkable achievements \cite{wang2020reinforcement, wang2021codet5, chen2021evaluating, li2022automating, zhu2022multilingual, guo2022unixcoder}, 
thanks to its powerful representation capability and capacity to uncover hidden patterns. 
At present,
a number of code intelligence tools leveraging deep learning, including Copilot and ChatGPT, have been developed and commercialized. 
These advanced tools enable software developers to write efficient and accurate code with greater ease and speed.

This survey focuses on the nl-to-code search task. 
Code search is one of the three main software engineering tasks that has remained active over the years \cite{watson2022systematic}. 
While code-to-code search falls within the research domain of code search, 
its role bears a closer affinity to code clone detection, 
thus rendering it outside the scope of this survey. 
The goal of nl-to-code search (henceforth referred to as code search) is to 
retrieve code fragments that match 
developers' natural language queries from a large code corpus \cite{stolee2014solving}.
Software development is a creative work that requires software developers to write code that meets product requirements.
During software development, 
developers often use search engines to search for code-related information \cite{shuai2020improving}, 
such as reusable code snippets, API understanding, and examples of specific functionality. 
They seek high-quality open source code for reference or reuse, 
which enhances the productivity of development \cite{gharehyazie2017some}. 
Developers typically use general search engines such as Google, Bing, and Baidu to search for target codes. 
However, these general search engines usually search in the full database and are not specifically designed for code-related information. 
Consequently, they are inefficient on the code search task and the search results are not satisfactory.

\textbf{The transformative impact of deep learning on code search.} 
Early code search engines primarily utilized Information Retrieval (IR) technology. 
They treated source code as text, 
and searched for the target code by matching keywords in query and code \cite{chatterjee2009sniff,mcmillan2011portfolio,hill2014nl,nie2016query}.
These methods mainly rely on the textual similarity between code and query. 
However, 
this approach has 
drawbacks since programming languages and natural languages differ greatly from each other, making it difficult for IR-based methods to comprehend the semantics.
Fortunately, 
deep learning boasts an exceptional capability for extracting high-level semantic representations. 
As a result, 
compared to conventional IR-based approaches, code search has been improved significantly since deep learning incorporation \cite{gu2018deep}. 
Deep learning has received great attention in recent years in code search research. 
Code search is one of the first software engineering tasks to use deep learning techniques \cite{watson2022systematic}.
Hence, this paper primarily investigates code search methods based on deep learning.
We provide a taxonomy which groups the current series of work into three categories:
various techniques are employed to enhance the semantics of the query text 
for a more precise search intent \cite{sirres2018augmenting, rahman2018effective, rahman2019automatic, rahman2019supporting, zhang2017expanding, 
liu2019neural, cao2021automated, huang2019deep, wu2019code, wang2022enriching, eberhart2022generating, sivaraman2019active};  
sequence and graph-based deep learning technologies are introduced to model code representation and enhance code comprehension
\cite{cambronero2019deep, feng2020codebert, ling2020adaptive, huang2021cosqa, lachaux2021dobf, du2021single, 
salza2022effectiveness, wang2022bridging, chai2022cross, li2022coderetriever, li2022soft, li2022exploring, 
guo2020graphcodebert, gu2021multimodal, xu2021two, wang2021syncobert, wang2022code, niu2022spt, guo2022unixcoder, 
gu2018deep, sachdev2018retrieval, wan2019multi, zeng2023degraphcs, ling2021deep, sun2022code, liu2023graphsearchnet, 
yao2019coacor, haldar2020multi, wang2020trans, zhao2020adversarial, ye2020leveraging, gu2021cradle, cheng2022csrs, 
arakelyan2022ns3, hu2022cssam, han2022towards, ma2023mulcs};
more efficient training techniques are introduced or proposed, making the training of large models a great success 
\cite{feng2020codebert, guo2020graphcodebert, wang2021syncobert, guo2022unixcoder, wang2022code, niu2022spt, lachaux2021dobf, 
li2022soft, huang2021cosqa, gu2018deep, wan2019multi, zeng2023degraphcs, ling2021deep, sun2022code, xu2021two, ling2020adaptive, 
liu2023graphsearchnet, li2022coderetriever, li2022exploring, chen2018neural, chai2022cross, wang2022bridging, 
han2022towards, yao2019coacor, haldar2020multi, wang2020trans, zhao2020adversarial, ye2020leveraging, 
gu2021cradle, bui2021self, cheng2022csrs, arakelyan2022ns3, park2023contrastive, hu2022cssam, ma2023mulcs, hu2023revisiting}.

\textbf{Significance of this review.} 
With numerous deep learning-based code search works published, 
especially the advent of pre-training technology in recent years, 
the field of code search has entered a new era. 
Despite the existence of several studies in the field of deep learning and code search, 
there has not been a thorough and organized examination of the correlation between them. 
As a result, we undertake a study of recent deep learning research in code search, 
develop a new taxonomy and provide insights into the future research opportunities from our discoveries.
This review serves to fill this gap. 
We believe that this review will be useful for both relevant academic researchers and industrial practitioners  
as follows:
\begin{enumerate}[(1)]
  \item Individuals who are new to the field of code search and 
  wish to gain a foundational understanding through a review.
  \item Individuals who already have some understanding of the field of code search, 
  but lack 
  an organized summary and classification, 
  and are unable to establish a 
  coherent knowledge structure for the field.
  \item Individuals who are interested in learning about the latest advancements and 
  state-of-the-art methods in the field of deep code search.
  \item Individuals who aim to develop code search tools in the industry.
  \item Individuals who are currently seeking a research direction in the field of deep code search.
\end{enumerate}

\textbf{Differences from prior reviews.}
At present, there have been several surveys on code search, 
but their focus is different from ours.
Khalifa \cite{khalifa2019semantic} discusses existing code search techniques, 
focusing on IR-based and deep-learning-based approaches, 
but only covers 5 relevant papers.
Liu et al. \cite{liu2021opportunities} focus on publication trends, 
application scenarios, and evaluation metrics of code search.
Grazia and Pradel \cite{grazia2022code} provide a review of the historical development of code search and 
covers the various stages of the code search process, 
including query formulation, query processing, indexing, and ranking.
However, the review lacks theoretical analysis and in-depth summaries, 
especially in terms of the relationship between query and code semantic matching models.
Different from the previous reviews, 
we concentrate on code search technology based on deep learning. 
We collect and organize high-quality conference/journal papers published in the past 5 years 
and propose a new taxonomy.
In this survey, we have discussed the following research questions (RQs):
\begin{itemize}
  \item \textbf{RQ1.} \textit{How to accurately capture the user's query intent?}
  \item \textbf{RQ2.} \textit{How to enhance the semantic understanding of code?}
  \item \textbf{RQ3.} \textit{How to train a deep code search model?}
  \item \textbf{RQ4.} \textit{How to evaluate a code search model?}
  \item \textbf{RQ5.} \textit{What are some potential avenues for future research in this promising field?}
\end{itemize}

\textbf{Literatures collection process.}
Our aim is to equip novice researchers and non-experts with a quick and systematic grasp of the latest code search techniques and motivate their future research. 
Considering that Gu et al. \cite{gu2018deep} were the pioneers in applying deep neural networks to code search tasks in 2018, 
we establish 2018 as our reference point and adhere to the guidelines proposed by Kitchenham and Charters \cite{keele2007guidelines} as well as Petersen et al \cite{petersen2015guidelines}. 
This allows us to perform a comprehensive examination of the relevant literature spanning the last six years (from 2018 to the present). 
Specifically, 
we identified a set of search strings, namely ``code search'' and ``code retrieval'', 
from the existing literature on code search that we are already known.
Subsequently, 
capitalizing on our emphasis on the precise technological domain of deep learning, 
we broadened our search strings to encompass the advanced concepts of ``deep code search'' and ``deep code retrieval''. 
Based on the aforementioned four search strings, 
we conducted initial manual searches on Google Scholar, DBLP, ACM Digital Library, and Papers With Code, 
resulting in 693 relevant papers. 
After removing duplicate papers, we identified a total of 431 candidate papers. 
Our comprehensive search was finalized on June 12, 2023, encompassing research published prior to this date. 
Subsequently, 
we meticulously applied a set of well-defined inclusion and exclusion criteria to identify the most relevant literature that aligns with our research topic:
\begin{itemize}
  \item[\Checkmark] Complete research papers published in top-ranked international academic conferences or journals. 
  Specifically, venues ranked A* or A in the CORE ranking: \href{https://www.core.edu.au/conference-portal}{http://portal.core.edu.au}.
  \item[\Checkmark] Significantly impactful papers from workshops, arXiv, and lower-ranked venues. 
  Specifically, these papers have received no less than 15 citations and are highly relevant to the subject of our investigation. 
  \item[\Checkmark] The paper must include a discussion or evaluation of the model on the task of natural language code search. 
  \item[\XSolidBrush] The preprints of accepted papers are excluded. 
  \item[\XSolidBrush] Journal papers that are extensions of conference proceedings are discarded.
  \item[\XSolidBrush] Non-deep learning-based code search papers are ruled out.
  \item[\XSolidBrush] Papers focusing on code-to-code search are excluded. 
  \item[\XSolidBrush] Papers that focus on utilizing code search models to assist other code intelligence tasks, 
  such as code generation and code repair, are discarded.
\end{itemize}

\begin{table}[H]
  \caption{\textbf{Selection of relevant papers.}}
  \centering
  \begin{tabular}{cc}
  \toprule
  Process & Number of papers \\
  \midrule
  Total papers retrieved & 693 \\
  \midrule
  After removing duplicate papers & 431 \\
  \midrule
  After excluding based on title, abstract, and keywords & 97 \\
  \midrule
  After excluding based on the full text & 64 \\
  \bottomrule
  \end{tabular}
\label{table:selection_of_papers}
\end{table}

\begin{table}[H]
  \caption{\textbf{Top publication venues with at least two deep code search papers.}}
  \centering
  \resizebox{\textwidth}{!}{
  \begin{tabular}{cccc}
  \toprule
  Publication venue types & Short Name & Full Name & Number of papers \\
  \midrule
  \multirow{12}*{Conference} & ICSE & International Conference on Software Engineering & 9 \\
                             & EMNLP & Conference on Empirical Methods in Natural Language Processing & 6 \\
                             & SANER & IEEE International Conference on Software Analysis, Evolution, and Reengineering & 5 \\
                             & ACL & Annual Meeting of the Association for Computational Linguistics & 4 \\
                             & FSE/ESEC & ACM Joint European Software Engineering Conference and Symposium on the Foundations of Software Engineering & 3 \\
                             & ASE & International Conference on Automated Software Engineering & 3 \\
                             & TSE & IEEE Transactions on Software Engineering & 3 \\
                             & NeurIPS & Conference on Neural Information Processing Systems & 3 \\
                             & ICPC & IEEE International Conference on Program Comprehension & 3 \\
                             & WWW & World Wide Web & 3 \\
                             & PLDI & ACM SIGPLAN Conference on Programming Language Design and Implementation & 2 \\
                             & ICSME & International Conference on Software Maintenance and Evolution & 2 \\
  \midrule
  \multirow{2}*{Journal}     & TOSEM & ACM Transactions on Software Engineering and Methodology & 2 \\
                             & - & Neural Networks & 2 \\
  \midrule  
  Total & - & - & 50 \\       
  \bottomrule
  \end{tabular}
  }
\label{table:publication_venues}
\end{table}

\begin{figure}[htbp]
  \centering
  \includegraphics[width=0.5\linewidth]{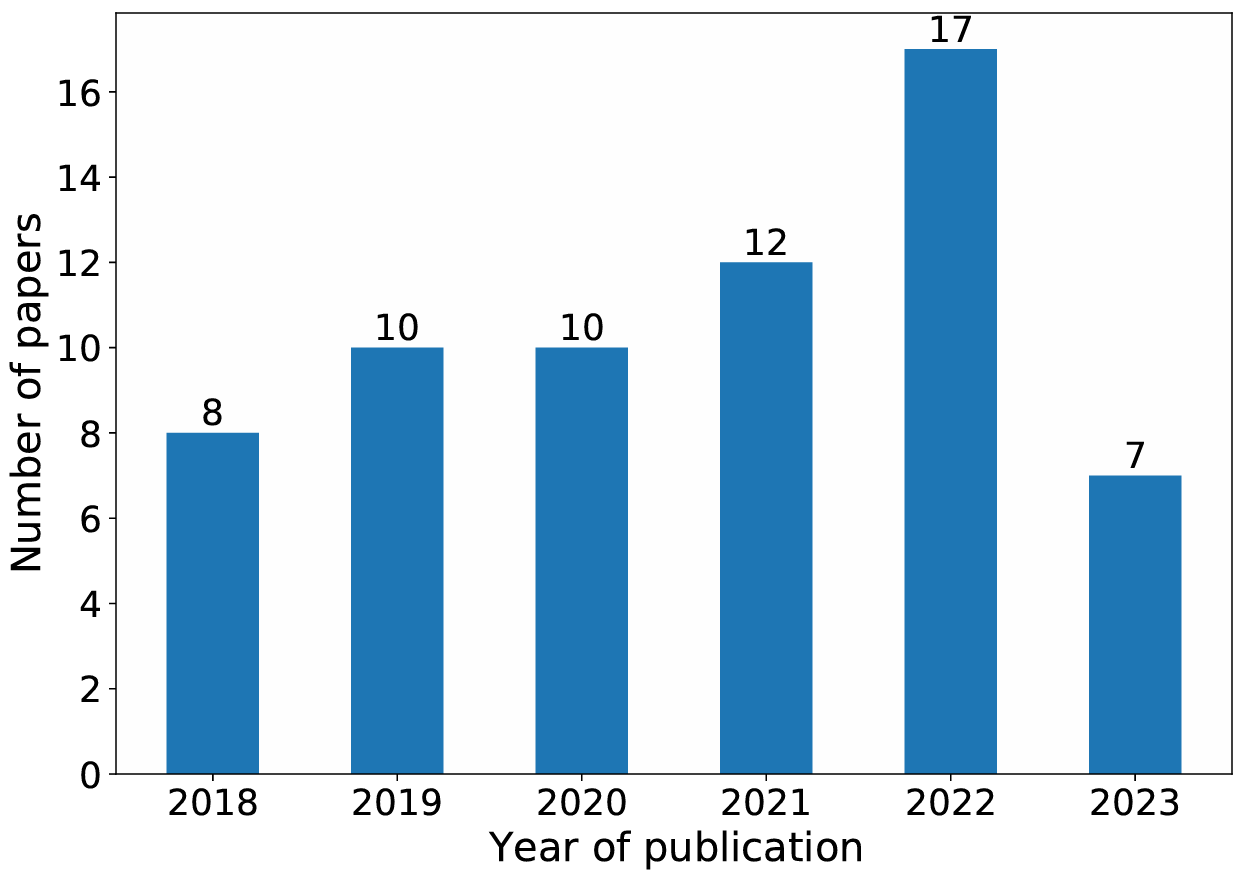}
  \caption{Number of publications per year.}
  \label{year_of_publication}
\end{figure}

As presented in Table \ref{table:selection_of_papers}, 
through a thorough analysis of the titles, abstracts, and keywords of the papers, 
we successfully identified 97 papers. 
Subsequently, 
after carefully cross-checking the full texts of the remaining papers, 
we ultimately obtained a collection of 64 papers on code search leveraging deep learning techniques. 
These papers primarily encompass international top conferences and journals in the domains of Software Engineering, Content Retrieval, and Artificial Intelligence.
Table \ref{table:publication_venues} presents the esteemed publication venues have published at least two deep code search papers. 
These venues collectively feature 50 papers, constituting 78.1\% of the entire obtained papers. 
Notably, conference proceedings accounted for 92\% of the published works. 
Among these 14 venues, 
we observed that the top five most popular conferences are ICSE, EMNLP, SANER, ACL, and FSE/ESEC, 
while the top two preferred journals are TOSEM and Neural Networks. 
Figure \ref{year_of_publication} showcases a compelling upward trajectory in the annual publication quantity of relevant literature following the integration of deep learning techniques into the field of code search. 
This notable trend serves as a testament to the escalating attention and burgeoning interest surrounding deep code search.

\textbf{Our contributions}. The main contributions of this paper are as follows:
\begin{enumerate}[(1)]
  \item New taxonomy. 
  We review 64 related works on deep code search published until June 12, 2023, 
  and propose a new taxonomy which groups code search based on deep learning into three categories: 
  query semantics modeling, code semantics modeling, and matching modeling.
  \item Comprehensive review. 
  We present a comprehensive overview of deep learning techniques in code search, 
  offering in-depth descriptions of representative models, thorough comparisons, and summarized algorithms. 
  \item Abundant resources. 
  We gather a vast array of code search resources, 
  including large-scale training corpus, benchmark datasets, and evaluation metrics suitable for various scenarios, 
  to aid in comparing deep code search models. 
  \item Future directions. 
  We highlight the limitations of current deep code search methods 
  and suggest various valuable research directions, 
  aiming to spur further exploration in this area. 
\end{enumerate}

The rest of this paper is organized as follows.
Section 2 introduces the deep learning-based code search framework and 
defines the core problem of code search. 
Section 3 covers methods for enriching the semantics of natural language text queries.
Section 4 categorizes various code representation and vectorization techniques.
Section 5 outlines the training methods and objectives of models.
Section 6 summarizes commonly used datasets and evaluation metrics in this field.
Section 7 highlights potential future research directions in this promising field. 
Section 8 provides a summary of this survey.

\section{Deep Learning-Based Code Search Framework}
Deep learning has seen rapid development in the field of software engineering in recent years, 
with code search being one of the most successful applications. 
Compared with traditional methods, 
deep code search leverages the powerful ability of deep neural network to 
extract hidden features from data, 
generating semantic representation of natural language and code 
for improved performance.

\begin{figure}[htbp]
  \centering
  \includegraphics[width=0.7\linewidth]{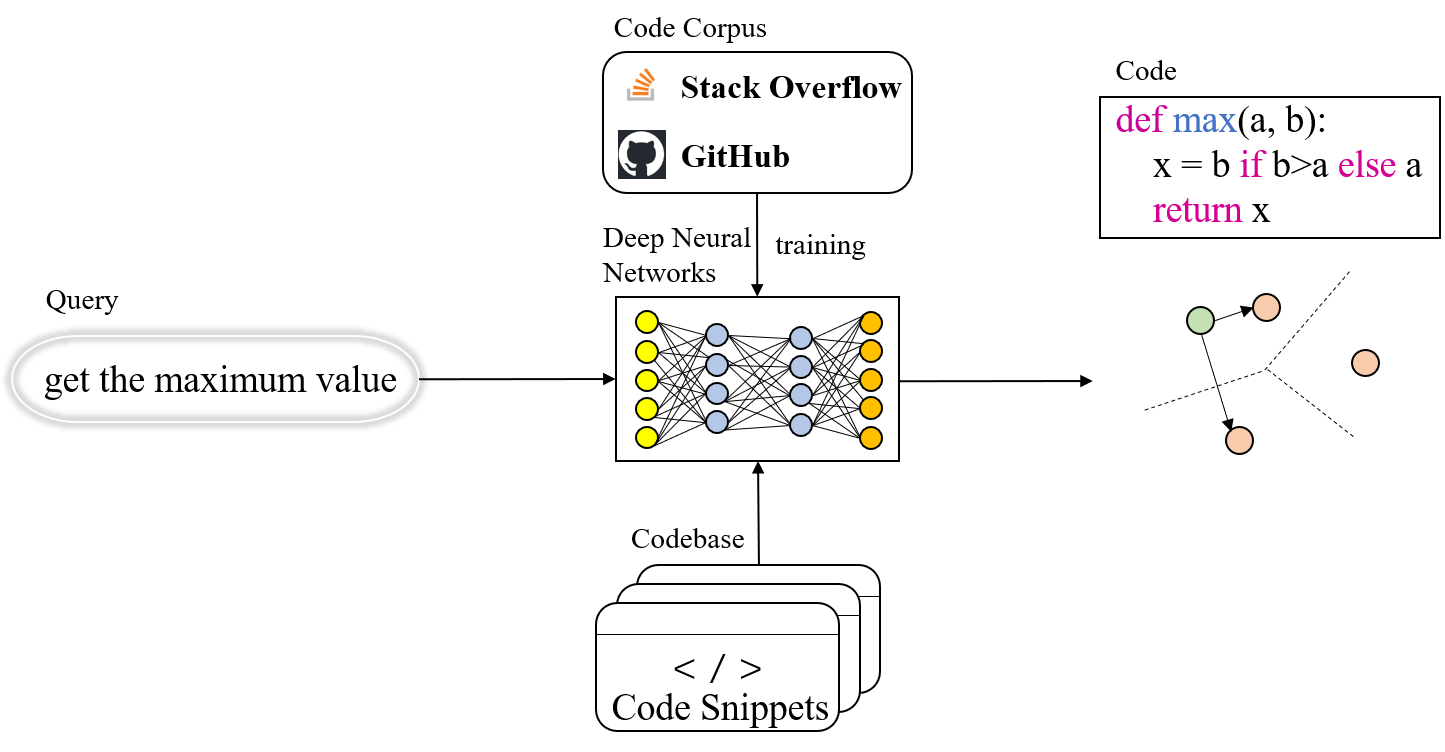}
  \caption{Code search framework based on deep learning.}
  \label{framework}
\end{figure}

Figure \ref{framework} illustrates the overall framework of deep learning-based code search.
The framework consists primarily of three components: 
encoding the query, encoding the code, and measuring their similarity. 
During training, 
a bimodal deep neural network is trained on a large parallel corpus of code and natural language to 
learn how to encode query and code into high-dimensional vectors, $ \textbf{q} $ and $ \textbf{c} $. 
Subsequently, the similarity between the query vector $ \textbf{q} $ and the code vector $ \textbf{c} $ in a high-dimensional vector space is calculated, 
such as Cosine similarity $ s (\textbf{q}, \textbf{c})=\frac{\textbf{q}^T \cdot \textbf{c}}{\|\textbf{q}\| \cdot\|\textbf{c}\|} $ or Euclidean distance $s(\boldsymbol{q}, \boldsymbol{c})=\sqrt{\sum_{i=1}^n\left(q_i-c_i\right)^2}$. 
Among these options, Cosine similarity stands out as the widely preferred calculation method 
\cite{gu2018deep, wan2019multi, ling2020adaptive, ling2021deep, xu2021two, li2022coderetriever, zeng2023degraphcs, hu2023revisiting}.

To improve code search accuracy, 
the model should make query representation similar to correct code representation
while distinct from incorrect code representation. 
Hence, training instances are typically structured as triplets $\left\langle q, c^{+}, c^{-}\right\rangle$, 
and the model is trained to minimize the triplet sorting loss $ \mathcal{L}\left(q, c^{+}, c^{-}\right)=\max \left(0, \delta-s \left(q, c^{+}\right)+s \left(q, c^{-}\right)\right) $, 
where $ q $ denotes the natural language query text, 
$ c^{+} $ denotes correct code fragment, 
$ c^{-} $ is a randomly selected negative sample code fragment from the rest of the code corpus, 
and $ \delta $ denotes the margin that ensures $ c^{+} $ is closer to $ q $ than $ c^{-} $ in the vector space.

To enhance the response efficiency of the online code search service, 
the trained model typically pre-calculates the vector representation of 
the candidate code fragments in the codebase. 
Upon receiving a query, 
the code search engine computes the vector representation of the query, 
traverses the codebase to compute the 
similarity between the query 
and each code fragment in the semantic space, 
and finally returns the top $k$ code fragments based on the relevance scores.

\begin{figure}[htbp]
  \centering
  \includegraphics[width=0.8\linewidth]{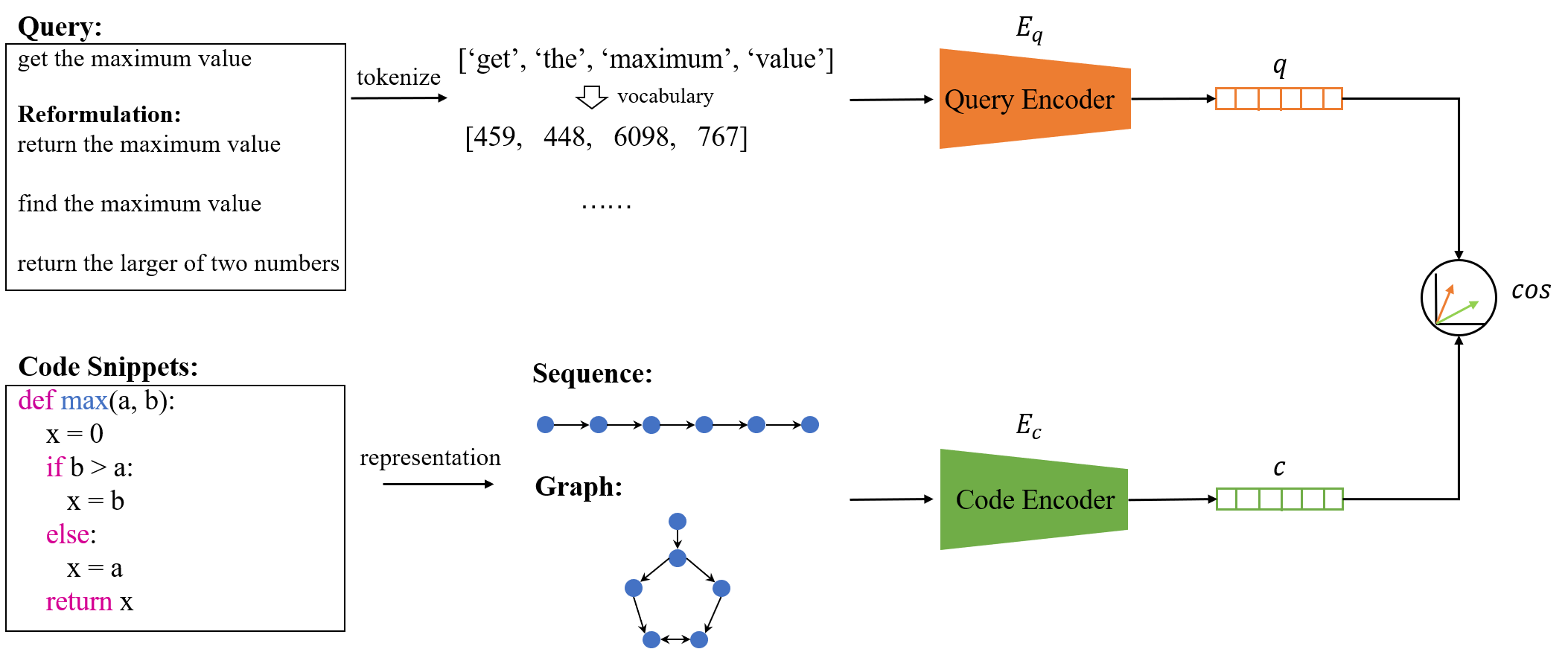}
  \caption{Overview of techniques for deep code search.}
  \label{overview}
\end{figure}

As shown in Figure \ref{overview}, 
deep learning-based code search engines mainly deal with two types of input: 
natural language query and source code. 
Deep learning techniques are widely employed to learn the semantic encoding of both inputs, 
especially the deep semantic information of source code. 
Following this overview, 
we want to categorize, analyze, and summarize 64 papers that are highly relevant to date and deep code search. 
To accomplish this objective, we undertook an exploration of five research questions (RQs):
\begin{itemize}
  \item \textbf{RQ1.} \textit{How to accurately capture the user's query intent?} 
To achieve the desired level of search result quality, 
it is crucial for the model to possess a good understanding of the user's query intent. 
The goal of this RQ is to investigate how to model the representation of queries and enhance the semantic representation of query, 
thereby assisting the model in accurately capturing the user's query intent.
  \item \textbf{RQ2.} \textit{How to enhance the semantic understanding of code?} 
The syntax of programming languages differs from natural language, giving rise to a semantic gap between them. 
This RQ investigates through multiple perspectives, encompassing the representation and vectorization of code, 
as well as diminishing the semantic gap between code and natural language through interaction.
  \item \textbf{RQ3.} \textit{How to train a deep code search model?} 
This RQ aims to investigate how to pretrain code intelligence large models 
and how to fine-tune code search models.
  \item \textbf{RQ4.} \textit{How to evaluate a code search model?} 
The goal of this RQ is to analyze code search datasets, evaluation metrics, 
and the selection of appropriate code search methods.
  \item \textbf{RQ5.} \textit{What are some potential avenues for future research in this promising field?} 
Through analysis of the aforementioned RQs, we have identified several limitations and challenges. 
Building upon this foundation, this RQ discusses twelve valuable future research directions in the field of deep code search. 
\end{itemize}
In the following sections, 
we will delve into a comprehensive and detailed discussion of each RQ.

\section{Query Representation and Enhancement (RQ1)}
Processing a user's input query is the initial step in conducting a code search task. 
Accurately comprehending the search intention within the query text is crucial in providing satisfying results to users. 
This section covers the query processing procedure from two perspectives: 
modeling the query representation and enhancing the semantic representation of the query.

\subsection{Query Representation}
The deep learning-based code search represents queries and code fragments as vectors in high-dimensional space by encoding them. 
The query encoder and the code encoder can use the same or different network architectures.
To derive the vector representation of the query, 
tokenization of the input query text is required. 
Tokenization involves reassembling continuous word sequences into token sequences based on specific criteria. 
As shown in Figure \ref{overview}, 
the developer inputs a query ``get the maximum value'', 
and the tokenizer processes it into a token sequence [``get'', ``the'', ``maximum'', ``value''], 
which is then looked up in the dictionary to get the id list [459, 448, 6098, 767]  
as the input for the query encoder.
The query encoder $ E_q $ is an embedding network that 
converts the developer-given query text into a $d$-dimensional vector representation.
To train the query encoder, 
existing deep learning-based code search methods have tried various neural network architectures, 
such as RNN \cite{gu2018deep}, LSTM \cite{zeng2023degraphcs}, GNN \cite{ling2021deep}, and transformer \cite{feng2020codebert}.
As collecting and annotating real (query, code) pairs is costly, 
comments are often treated as queries during training.

\subsection{Methods For Enhancing query Representation}
Developers often struggle to clearly articulate their target requirements when searching for code 
and prefer to use short initial queries with broad meanings, such as keywords splicing.
In this case, 
the code search engine is likely to return many irrelevant code fragments. 
As a result,
Developers have to waste time reviewing non-relevant results and adjusting queries.
To address this, 
several studies have focused on enhancing query semantics to improve code search efficiency 
by enabling the code search engine to expand or reconstruct the query automatically. 
Table \ref{table:enhancing_query} summarizes the different approaches along 
based on API or class name, based on co-occurring terms, based on commit versions, 
based on query generation, based on search logs, based on user feedback, and based on multiple perspectives. 
We will discuss them in detail below.

\begin{table}[H]
  \caption{\textbf{Overview of approaches for enhancing query representation.}}
  \centering
  \resizebox{\textwidth}{!}{
  \begin{tabular}{ccccccc}
  \toprule
  \multirow{2}{*}{Year} & \multirow{2}{*}{Work}            & \multicolumn{5}{c}{Based on} \\
  \cmidrule(lr){3-7}
   & &  API or class name & co-occurring terms & commit versions &  query generation & Others \\
  \midrule
  2018 & NLP2API \cite{rahman2018effective} & \Checkmark & & & & \\
  \midrule
  2018 & COCABU \cite{sirres2018augmenting} & \Checkmark & & & & \\
  \midrule
  2018 & Zhang et al. \cite{zhang2017expanding} & \Checkmark & & & & \\
  \midrule
  2019 & RACK \cite{rahman2019automatic} & & \Checkmark & & &  \\
  \midrule
  2019 & Rahman \cite{rahman2019supporting} & & \Checkmark & & & \Checkmark \\
  \midrule
  2019 & ALICE \cite{sivaraman2019active} & & & & & \Checkmark \\
  \midrule
  2019 & NQE \cite{liu2019neural} & \Checkmark & & & & \\
  \midrule
  2019 & QESC \cite{huang2019deep} & & & \Checkmark & & \\
  \midrule
  2019 & Wu and Yang \cite{wu2019code} & & & \Checkmark & &   \\
  \midrule
  2021 & SEQUER \cite{cao2021automated} & & & & & \Checkmark \\
  \midrule
  2022 & QueCos \cite{wang2022enriching} & & & & \Checkmark & \\
  \midrule
  2022 & ZaCQ \cite{eberhart2022generating} & & & & \Checkmark & \\
  \bottomrule
  \end{tabular}
  }
\label{table:enhancing_query}
\end{table}

\textbf{Based on API or class name.}
Enhancing natural language queries with semantically related identifiers 
has significant potential.
Researchers have leveraged source code knowledge from platforms 
such as Stack Overflow and Github to reformulate natural language queries.
The commonality among query reformulation methods is 
using similarity measures to compare the words or identifiers in the query to 
words or identifiers such as API and class names in source code \cite{sirres2018augmenting, rahman2018effective, rahman2019automatic, rahman2019supporting, zhang2017expanding}.
For instance, 
based on posts on Stack Overflow, Sirres et al. \cite{sirres2018augmenting} extend users' free-form queries 
with API methods or class names in code snippets marked as acceptable.
Rahman and Roy \cite{rahman2018effective} propose automatically identifying relevant or specific API classes 
from Stack Overflow to reformulate queries.
They first use pseudo-relevance feedback and term weighting algorithm 
to gather candidate API classes from Stack Overflow, 
and then apply the maximum likelihood estimation between the keywords in the query and API classes 
to rank them. 
Finally, the top-$k$ correlated classes are used to reformulate the query. 
Zhang et al. \cite{zhang2017expanding} apply a continuous bag-of-words model 
to learn the vector representation of API class names in the code corpus. 
These vectors are then utilized to calculate the semantic distance between the initial query and API class names. 
The most semantically relevant API class names are selected for query expansion.

\textbf{Based on co-occurring terms.}
A few approaches explore term co-occurrence relationships when reconstructing queries. 
These methods identify query keywords within structured entities like method and field signatures 
and then propose their co-occurring terms as candidates for query reformulation. 
For example,
Liu et al. \cite{liu2019neural} present a model called NQE that 
uses an encoder-decoder architecture to predict co-occurring keywords with query keywords in codebase. 
NQE takes an initial query as input to obtain the corresponding encoding, 
and then inputs this encoding into a Recurrent Neural Network (RNN) decoder as the initial state.
By collecting the output at each time step, 
NQE generates a sequence of method names.
Finally, an output set of expanded keywords is obtained by 
splitting each method name in the sequence. 

\textbf{Based on commit versions.}
The code retrieved via code search engines may require modification prior to use, 
such as for compatibility with different API versions. 
A few approaches take this potential code change into account when expanding queries, 
freeing developers from manually modifying codes.
Huang et al. \cite{huang2019deep} analyze the changes made to codes 
through extracted sequences from GitHub commits to 
determine what modifications were made and the reasons. 
This aids in inferring fine-grained changes to queries. 
They expand the query with relevant terms and eliminate irrelevant terms 
to minimize the negative effects of excessive query expansion.
Wu and Yang \cite{wu2019code} analyze recurrent code changes within the versions history of open source projects.
For example, 
if token ``$a$'' in the code snippet is frequently modified to token ``$b$'', 
then when the target code contains token ``$a$'', the token ``$b$'' can be used to expand the initial query.
By expanding the query in this way, 
the search engine can retrieve the updated code snippets, 
avoiding developers manually updating query.

% \textbf{Based on search logs.}
% In addition to utilizing identifiers such as APIs and class names in source code, 
% there are also 
% methods to explore the search logs from Stack Overflow.
% Cao et al. \cite{cao2021automated} provide a unique insight into query reformulation patterns 
% based on large-scale real search logs from Stack Overflow.
% They build a large-scale query reconstruction corpus, 
% incorporating both original queries and their corresponding reconstructed queries, 
% to train the model. 
% When given a user query, 
% the trained model automatically generates a list of 
% candidate reconstructed queries for selection.

\textbf{Based on query generation.}
Besides extracting relevant identifiers or keywords from codebases 
to expand queries, 
a few approaches aim to generate queries with richer semantics.
Wang et al. \cite{wang2022enriching} argue that many existing deep learning methods in the field 
overlook the knowledge gap between query and code description. 
They note that queries are typically shorter than the corresponding code descriptions.  
Moreover, using (code, description) pairs as training data may not generalize well to real-world user queries. 
To address these issues, they propose the model QueCos. 
QueCos employs LSTM with an attention mechanism as its architecture 
and uses reinforcement learning to capture the key semantics of a given query. 
It then generates a semantically enhanced query, 
and blends the results of the initial query and the semantically enhanced query 
using a mix-ranking technique to prioritize relevant code snippets. 
To solve the problem that the ground-truth answer
does not have a high rank, 
Eberhart and McMillan \cite{eberhart2022generating} propose the model ZaCQ 
to generate questions that clarify the developers' search intent.
When given a code search query and the top-$k$ most relevant results, 
ZaCQ identifies ambiguous aspects in the query, 
extracts task information from attributes such as function names and comments, 
and generates a targeted clarifying question for unclear requirements. 
Eventually, it employs the feedback association algorithm to boost the ranking of relevant results.

\textbf{Others.}
Rahman \cite{rahman2019supporting} reformulates the query from multiple perspectives to further enrich its semantics.
He gathers search keywords from texts and source codes, 
structured entities from bug reports and documents, 
and relevant API classes from Stack Overflow Q\&A posts. 
After that, he uses this information to reformulate queries 
and improve code search performance with appropriate term weighting and context awareness. 
To explore the search logs from Stack Overflow, 
Cao et al. \cite{cao2021automated} provide a unique insight into query reformulation patterns 
based on large-scale real search logs from Stack Overflow.
They build a large-scale query reconstruction corpus, 
incorporating both original queries and their corresponding reconstructed queries, 
to train the model. 
When given a user query, 
the trained model automatically generates a list of 
candidate reconstructed queries for selection. 
User feedback can also be actively included to help users refine query. 
Sivaraman et al. \cite{sivaraman2019active} leverage user feedback to label whether the returned samples are desired or not desired, 
and then extract the logical information of the code from these positive and negative samples to reconstruct the query.

\subsection{Summary} 
Current code search engines struggle with understanding natural language queries 
and only return relevant code fragments if the query includes specific identifiers like class and function names. 
However, 
developers may 
not know the relevant identifiers. 
To overcome this, 
data from sources such as Stack Overflow, GitHub, and codebases are fully mined 
to extract valuable information, 
such as code elements and term tokens, 
based on API, co-occurrence relations, search logs, and commit information. 
This information is then 
added to natural language queries with appropriate term weighting, 
significantly improving the representation of the query.  
Moreover, 
techniques exist to produce a query with a more lucid search intent by leveraging the extracted information, 
thereby enhancing search efficiency and performance. 
However, semantically related terms may not always co-occur,  
and simple co-occurrence frequencies may not be sufficient for 
selecting appropriate search keywords for query reformulation. 
More importantly, 
the success of the extended query method also relies on the quality of the extracted words. 
The crowdsourced knowledge from Stack Overflow may contain noise 
and if the words are not extracted properly, 
it can lead to excessive query expansion and negatively impact search accuracy. 

\newtcolorbox{RQ1box}{colback=white, colframe = black}
\begin{RQ1box}
  \textbf{\textit{Summary of answers to RQ1: }}
  \begin{itemize}
    \item \textit{Information such as API or class name, co-occurring terms, commit versions, 
search logs and user feedback are often used to reformulate the query.}
    \item \textit{Reformulating query based on API or class name is the most popular method in the past 6 years.}
    \item \textit{When reconstructing the query, it is important to avoid introducing noisy words that could result in excessive query expansion.}
  \end{itemize}
\end{RQ1box}

\section{Code representation and vectorization (RQ2)}
The goal of a code search engine is to retrieve code fragments that match the query semantics from the codebase. 
To close the semantic gap between programming language and natural language, 
deepening the understanding of code semantics and fostering robust interaction between code and query are essential 
in addition to improving the accuracy of query intent. 
In this section, 
we discuss the code encoding model, 
examining it through the lenses of code representation and code vectorization. 
Additionally, 
we discuss the crucial aspect of the interaction between query and code, 
shedding light on its significance for the comprehensive semantic understanding of both elements.

\subsection{Code Representation}

\subsubsection{Preliminaries}
Source code is the original program text written by developers in a programming language, 
consisting of instructions and statements. 
It is typically compiled into machine language that a computer can understand and execute.
During the compilation process from source code to machine code, 
various intermediate representation forms are generated, 
such as Abstract Syntax Tree (AST), Data Flow Graph (DFG), Control Flow Graph (CFG), and LLVM Intermediate Representation (IR). 
In this process, 
the compiler automatically performs program analysis techniques, 
including lexical, syntactic, and semantic analysis, 
to verify the correctness of the source code.

\begin{figure}[htbp]
  \centering
  \includegraphics[width=0.8\linewidth]{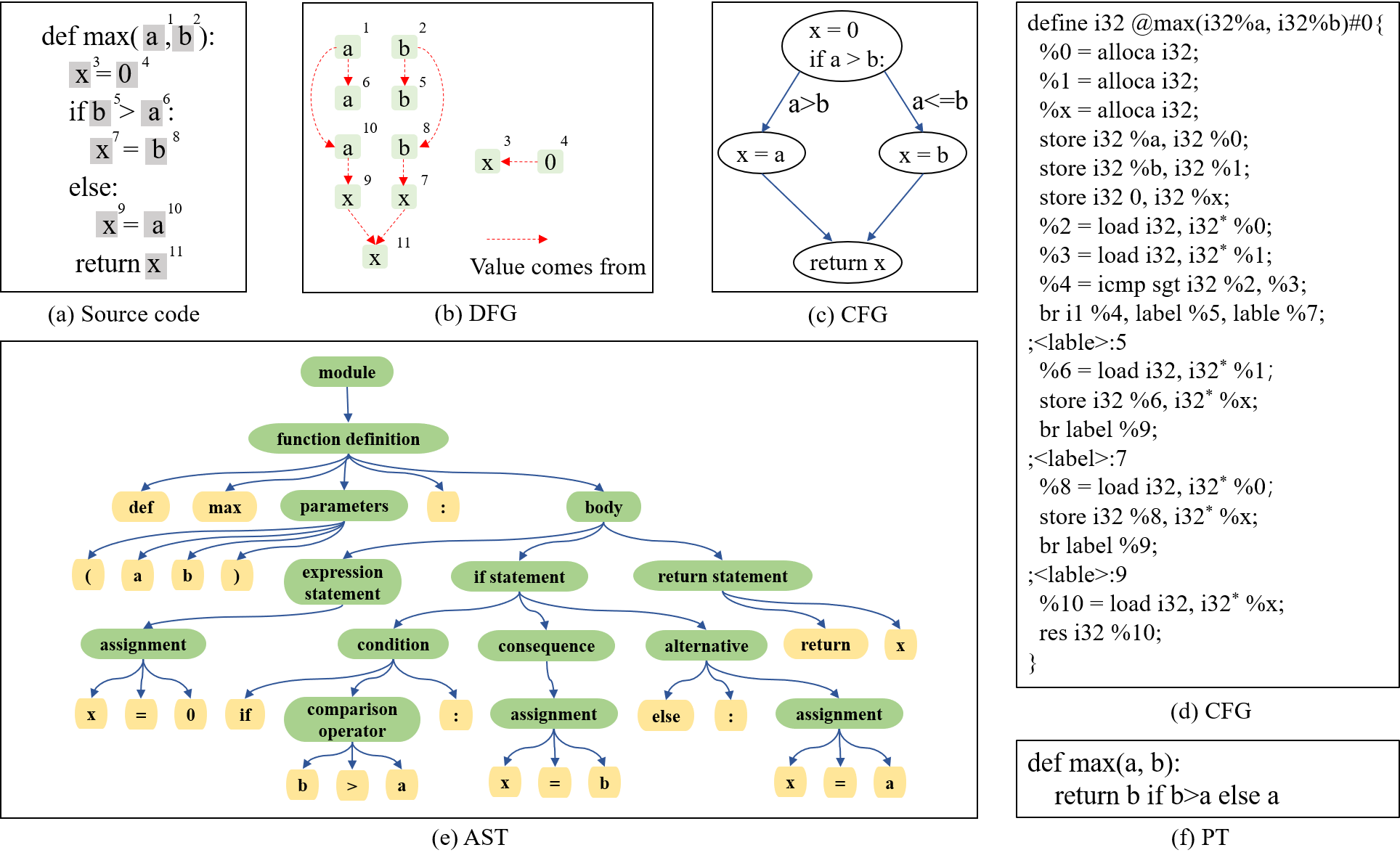}
  \caption{Multiple Modalities of Source Code.}
  \label{code_representation}
\end{figure}

\textbf{Abstract Syntax Tree (AST)} 
is a tree-like representation of the source code's syntax. 
It consists of leaf nodes, non-leaf nodes and the edges between them, 
reflecting the source code's rich syntax structure. 
For example, 
in Figure \ref{code_representation}(e), 
the assignment statement ``$ x=0 $'' is represented by a non-leaf node ``$assignment$''
and three leaf nodes ``$x$'', ``$=$'', and ``$0$'' connected by directed edges. 
A conditional jump statement like $ if-condition-else $ 
can be represented by a node with three branches.
Typically, the standard compiler tool $ tree-sitter $ \footnote{https://github.com/tree-sitter/tree-sitter}
is commonly used to parse source code into AST.

\textbf{Data Flow Graph (DFG)} 
represents the logical relationships of the code, 
the direction of data transmission and the process of logical transformation.  
DFG abstracts the dependencies between variables in the source code, 
where nodes represent variables and edges represent the source of each variable's value \cite{guo2020graphcodebert}. 
As shown in Figure \ref{code_representation}(b), 
the variable $x^{11}$ returned by the function has two sources 
which can either come from $a^{10}$ or $b^8$ based on the outcome of the $if$ condition. 
Evidently, 
adhering to naming conventions is not universal among software developers, 
and this can pose a challenge in comprehending variable semantics at times. 
Fortunately, 
DFG helps overcome the semantic deviation caused by inconsistent naming.

\textbf{Control Flow Graph (CFG)} 
is an abstract representation of a program's structure, 
composed of basic blocks and directed edges between them.  
Each directed edge reflects the execution order of two basic blocks. 
For example, 
as shown in Figure \ref{code_representation}(c), 
the outcome of an $if$ statement results in two different execution paths, 
namely ``$x=a$'' and ``$x=b$''. 
CFG represents the running sequence and logical relationship of the code, 
effectively reflecting the execution semantics of the program.

\textbf{Intermediate Representation (IR)}
is a structured and language-agnostic representation of the source code that captures the essential semantics and structure of the code, 
while abstracting away language-specific details. 
It serves as an intermediate step during compilation or interpretation and enables various optimizations and transformations to be performed on the code. 
For example, 
as shown in Figure \ref{code_representation}(d), 
the assignment statement ``$ x=0 $'' is represented by the IR instruction \verb| %x = alloca i32; store i32 0, i32 %x; |, 
that is, first allocate storage space for the temporary variable ``$x$'', 
and then store the constant ``$0$'' in it. 
IR serves as the foundation for building CFG and DFG.  
The common IR is shared by multiple objects 
and can realize the unified representation of different programming languages.

\textbf{Program Transformation (PT)}
denotes a code snippet that performs the same task as a given program.
It allows for the creation of numerous versions that fulfill identical functional requirements. 
For example, 
Figure \ref{code_representation}(f) is another representation 
of the functionality expressed in Figure \ref{code_representation}(a).

\subsubsection{Sequence}
The triumph of embedding technology in NLP has spurred researchers to investigate its potential use in code analysis. Presently, several studies treat source code as a straightforward text sequence and employ NLP techniques directly on code snippets.

\textbf{Source code Token Sequence (STS).}
Some approaches consider the tokenized code token sequence itself as the code representation 
and feed it into the model training process to learn the semantics of the code 
\cite{cambronero2019deep, yao2019coacor, zhao2020adversarial, ye2020leveraging, feng2020codebert, ling2020adaptive, huang2021cosqa, lachaux2021dobf, du2021single, 
salza2022effectiveness, arakelyan2022ns3, hu2022cssam, wang2022bridging, chai2022cross, li2022coderetriever, li2022soft, li2022exploring}.
For instance, 
based on the powerful learning ability of Transformer, 
Feng et al. \cite{feng2020codebert} build CodeBERT, the first large-scale program language pre-training model.
Inspired by BERT-based NLP,
they directly feed the token sequence of a code fragment into a multi-layer Transformer. 
This allows the model to incorporate context information and 
learn to identify tokens that are critical to the code's semantics. 
Nevertheless, 
obtaining all the tokens of the source code directly through tokenization ignores syntactic details, 
making it difficult to discern whether a term originates from a variable name or a method call. 
To this end, 
Du et al. \cite{du2021single} perform data augmentation on code fragments from three perspectives: 
code structure, variable names, and APIs. 
They construct structure-centric datasets, variable-centric datasets, and API-centric datasets, 
which are respectively fed into the model for training.
To assist the model in learning code features 
that are invariant across semantically equivalent programs, 
Wang et al. \cite{wang2022bridging} develop a semantic-preserving transformation for code snippets. 
The transformation preserves the code's semantics 
but changes its lexical appearance and syntactic structure. 
Specifically, 
they modify the control structure, APIs, and declarations of the code to 
enable the model to extract and learn relevant features, 
while preserving code semantics. 
Transformation-based approaches (such as generating semantically equivalent code fragments through variable renaming) 
often produce code with highly similar superficial forms, 
causing the model to focus on surface-level code structure rather than its semantic information. 
To mitigate this, 
Li et al. \cite{li2022soft} construct positive samples 
using code comments and abstract syntax tree subtrees,  
encouraging the model to capture semantic information.

\textbf{Multimodal Token Sequence (MTS).}
Code snippets have different information dimensions. 
A code token sequence only captures shallow features of source code, 
such as method names and code tokens, 
but ignores structural features like AST and CFG, 
which hold rich and well-defined source code semantics. 

Multimodal learning aims to build models that can process and aggregate information from multiple modalities. 
Recently, many studies have attempted to utilize program analysis techniques 
to capture the structural and syntactic representation of programs. 
These works capture multiple code modalities to form complementary code representations 
\cite{haldar2020multi, guo2020graphcodebert, gu2021cradle, gu2021multimodal, xu2021two, wang2021syncobert, wang2022code, han2022towards, niu2022spt, guo2022unixcoder}.
For instance, 
Xu et al. \cite{xu2021two} extract AST from the method body as a structural feature  
and parse a code fragment into a syntax tree. 
The nodes in the tree represent various code types, 
such as loop structures, conditional judgment structures, method calls, and variable declarations.
They traverse the AST through a breadth-first traversal strategy  
to obtain the AST node sequence, 
which they use for model training along with method names, API sequences, and code tokens.

To enhance the semantic suitability of the AST tree structure for code search,
Gu et al. \cite{gu2021multimodal} transform AST into Simplified Semantic Tree (SST). 
In contrast to AST, 
SST eliminates redundant nodes, enhances node labels, accentuates the semantic information of code snippets, 
and has broader applicability across diverse programming languages. 
Then they obtain tree-serialized representations from SST 
by sampling tree paths or traversing tree structures, 
which are used in multimodal learning to complement traditional source code token sequence representations. 
Similarly, Niu et al. \cite{niu2022spt} enrich the input representation of source code pre-trained models 
with simplified and linearized AST versions.
To facilitate the transformer in encoding AST, 
Guo et al. \cite{guo2022unixcoder} propose a method for lossless serialization of AST. 
They transform AST into a sequential structure that maintains all tree information 
and use it as input to enhance the code representation. 
To provide complementary information for program semantic understanding, 
Guo et al. \cite{guo2020graphcodebert} construct data flow graphs based on AST. 
Given a code fragment, 
they use a standard compiler to generate a AST, 
identify the variable sequence through the leaf nodes of the AST, 
and retain the source of each variable value to obtain the code's data flow information. 
Compared to AST, 
the data flow information is lighter and does not bring redundant structures, 
making the model more efficient. 

Wang et al. \cite{wang2022code} recognize that different views of code offer complementary semantics. 
They utilize the compiler to 
convert the program into multiple views, such as AST, CFG, and equivalent programs. 
AST provides the grammatical information of the code, 
CFG reveals the execution information of the code, 
and different variants of the same program offer the functional information of the code. 
They use a depth-first traversal to convert a AST into a sequence of AST tokens, 
and traverse a CFG along directed edges to parse it into a sequence of tokens. 
Finally, the model learns complementary information among multiple views 
under the framework of contrastive learning.

\begin{figure}[htbp]
  \centering
  \includegraphics[width=1\linewidth]{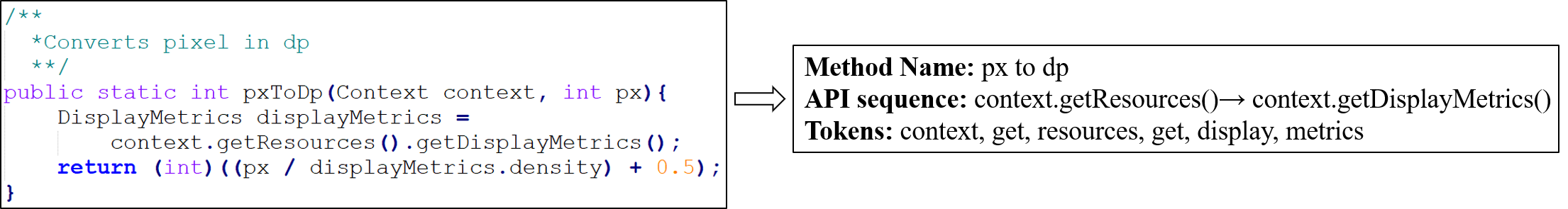}
  \caption{Extract feature tokens from code snippet.}
  \label{feature_tokens}
\end{figure}

\textbf{Feature Token Sequence (FTS).} 
Gu et al. \cite{gu2018deep} propose CODEnn, 
the first deep learning-based supervised model for code search tasks. 
Considering that the source code is not plain text, 
they represent it from three aspects: 
method name, API call sequence, and code tokens, 
as shown in Figure \ref{feature_tokens}. 
They encode each aspect separately and then 
combine them into a single vector to represent the entire code fragment.
Sachdev et al. \cite{sachdev2018retrieval} assume that source code tokens
contain enough natural language information. 
To represent code information at the method level, 
they construct a natural language document  
by extracting method names, method calls, enumerations, string constants, and comments. 
These methods generally follow the convention of not extracting variable names, which can vary from one software developer to another.  
Cheng and Kuang \cite{cheng2022csrs} employ a comprehensive approach to represent code fragments by utilizing method name, API sequence, and code tokens. 
They effectively convert each type of feature into its respective n-gram vector using the embedding layer. 
Eventually, 
they intelligently concatenate the three distinct feature vectors, 
resulting in the acquisition of the final code fragment feature matrix.

\subsubsection{Tree/Graph}
Converting natural graph structures such as AST or CFG into sequences can result in loss of key structural information. 
To address this, 
some approaches try to directly use the graph structure as the code representation. 
Wan et al. \cite{wan2019multi} regard CFG as a directed graph 
and use Gated Graph Neural Network (GGNN) to obtain the vector representation of CFG. 
They define a graph as $\mathcal{G}=\{\mathcal{V}, \mathcal{E}\}$, 
where $\mathcal{V}$ is a set of vertices $\left(v, \ell_v\right)$,   
$\mathcal{E}$ is a set of edges $\left(v_i, v_j, \ell_e\right)$,  
and $\ell_v$ and $\ell_e$ are the labels of vertices and edges, respectively. 
In the code retrieval scenario, 
each vertex represents a node in the CFG, 
and each edge signifies the control flow of the code. 

The code snippets 
with the same functionality may have different implementations, 
while the code snippets 
with different code semantics may have similar structural features. 
To achieve satisfactory results in code search, 
Zeng et al. \cite{zeng2023degraphcs} aim to find a more precise representation for the source code. 
They address the limitations of structural feature-based code representation methods by 
incorporating data and control dependencies. 
This allows semantically similar codes to have similar representations. 
To achieve this, 
they analyze various types of LLVM IR instructions, 
integrate data dependencies and control dependencies into a graph, 
and build a Variable-based Flow Graph (VFG). 
The nodes in this graph can be variables, opcode, or tag identifier. 
Considering that excessive information can obstruct the model from 
learning the fine-grained relationship between source code and query, 
they optimize VFG to minimize noise during training. 
Finally, 
they feed the optimized graph into a GGNN with an attention mechanism 
to learn the vector representation of the code. 

With the aim of aligning the semantics of query and code, 
Ling et al. \cite{ling2021deep} represent query text and source code as a unified graph structure. 
The program graph is generated from the AST of a code fragment.
The AST is represented as a tuple 
$\langle\mathbb{N}, \mathbb{T}, \mathbb{X}, \Delta, \phi, s\rangle$, 
where $\mathbb{N}$ denotes a set of non-terminal nodes, 
$\mathbb{T}$ denotes a set of terminal nodes, 
$\mathbb{X}$ represents a set of values,  
$\Delta: \mathbb{N} \rightarrow(\mathbb{N} \cup \mathbb{T})^*$ is a function that maps a non-terminal node to its child nodes, 
$\phi: \mathbb{T} \rightarrow \mathbb{X}$ is a function that maps a terminal node to an associated value, 
and $s$ represents the root node. 
Specifically, 
the program graph is composed of syntactic nodes 
(terminal and non-terminal nodes in AST) 
and grammatical tokens 
(the corresponding values of terminal nodes in the source code). 
This graph uses various types of edges
to model syntactic and semantic relationships between nodes and tokens, 
including Child edges to link syntactic nodes in the AST, 
NextToken edges to connect each syntactic token to its subsequent in the original code, 
and LastLexicalUse edges to associate identifiers with their nearest lexical usage in the code.

\subsection{Code Vectorization} 
Once the code representation is obtained, 
we can embed it in a vector space. 
The choice of network architecture for encoding depends on the type of 
code representation (such as source code tokens or AST) and data structure (such as AST sequences or graphs). 
The deep neural network exhibits its excellent performance in this process. 
In this section, 
the code embedding network based on deep learning is divided into five categories: 
Word Embedding, Recurrent Neural Network (RNN), Graph Neural Network (GNN), Transformer, and mixed mode. 
These will be discussed in details below.

\subsubsection{Based on Word Embedding} 
code2vec \cite{alon2019code2vec} is a seminal work in the field of code understanding, 
which realizes the representation of a code snippet as a single fixed-length code vector.  
These vectors serve as powerful tools to predict the semantic properties of the code snippet. 
Prior to this, 
the traditional word embedding algorithms include Word2Vec and GloVe. 
Word2Vec employs a two-layer dense neural network 
to calculate the vector representation of words. 
It can be trained unsupervisedly on large corpora, 
ensuring that words with common context are closely located in the vector space. 
FastText, a variant of Word2Vec, enhances word vectors with subword information 
and was once popular among researchers for code vectorization \cite{sachdev2018retrieval, cambronero2019deep, ling2020adaptive}. 
Saksham et al. \cite{sachdev2018retrieval} extract five types of information from each method:  
namely method name, method call, enumeration, string constant, and comment, 
to obtain the token sequence $ c=\left\{c_1, \ldots, c_n\right\}$ in the source code. 
Then they utilize FastText to calculate the vector representation 
of the extracted tokens sequence 
and obtain the embedding matrix $T \in \mathbb{R}^{\left|V_c\right| \times d}$, 
where $\left|V_c\right|$ is the size of the extracted code vocabulary,  
$d$ is the dimension of selected token embedding, 
and the $k$-th row in $T$ is the vector representation of the $k$-th extracted word in $V_c$, 
corresponding to the set of embedding vectors $\left\{T\left[c_1\right], \ldots, T\left[c_n\right]\right\}$ of the code. 
To combine a set of embedding vectors of code tokens into a single vector, 
they apply a weighted sum using TF-IDF weights. 
TF-IDF weights are designed to enhance the importance of tokens that frequently appear in code snippets, 
and reduce the weight of tokens that are commonly used across all codebases. 
When searching for code, 
they use $T$ to calculate the embedding vector $\left\{T\left[q_1\right], \ldots, T\left[q_m\right]\right\}$ 
of a given query $q=\left\{q_1, \ldots, q_m\right\}$, 
take its average value to obtain the final query vector $e_q$, 
and then use FAISS to calculate the distance between $e_c$ and $e_q$,  
representing the relevance of the code fragment to the given query. 
In the second year, 
they further optimize the code vectorization process by using supervised learning to update the token matrix $T$ obtained from FastText 
and replacing the TF-IDF weight of the merged token embedding with an attention-based scheme \cite{cambronero2019deep}. 
The attention weight $a_c \in \mathbb{R}^d$ is a $d$-dimensional vector learned during training.  
For the code token embedding vector $\left\{T\left[c_1\right], \ldots, T\left[c_n\right]\right\}$, 
the attention weight of each token is calculated as follows: 
\begin{equation}
\label{equa3}
\alpha_i=\frac{\exp \left(a_c \cdot T\left[c_i\right]^{\top}\right)}{\sum_{i=1}^n \exp \left(a_c \cdot T\left[c_i\right]^{\top}\right)}.
\end{equation}
Finally, 
the code's embedding vector is obtained by weighted summation of the attention weight $\alpha_i$: 
\begin{equation}
\label{equa4}
e_c=\sum_{i=1}^n \alpha_i T\left[c_i\right].
\end{equation}

\subsubsection{Based on Recurrent Neural Networks}
Recurrent Neural Networks (RNNs) are widely used for embedding sequences in natural language processing. 
Inspired by natural language processing techniques, 
some researchers have attempted to use RNN to encode code token sequences. 
For instance, 
Gu et al. \cite{gu2018deep} extract method names, API call sequences, and tokens from code snippets, 
represented as $C=[M, A, \Gamma]$, 
where $M=w_1, \ldots, w_{N_M}$ is a sequence of $N_M$ tokens obtained by dividing the method name with camel case, 
$A=a_1, \ldots, a_{N_A}$ is a sequence of $N_A$ consecutive API calls, 
and $\Gamma=\left\{\tau_1, \ldots, \tau_{N_{\Gamma}}\right\}$ is a collection of tokens in code fragments. 
They then encode the method name sequence and the API sequence separately using an RNN with a max-pooling layer. 
Since the tokens do not have a strict order in the source code, 
they are encoded using a multi-layer perceptron. 
Finally, the three vectors are combined into a single vector $c$, representing the entire code snippet: 
\begin{equation}
\label{equa5}
c={LSTM}_1(M)+{LSTM}_2(A)+{MLP}(\Gamma).
\end{equation}

Sun et al. \cite{sun2022code} translate the code into a natural language description to obtain a translation, 
and apply the LSTM architecture to build a translation encoder. 
They use word embedding to map words in the translation sequence, $s=w_1, \cdots, w_{N^s}$, to vector representations $\boldsymbol{w}_i=\psi\left(w_i\right)$.
These vectors are then arranged in an embedding matrix $E \in \mathbb{R}^{n \times m}$, 
where $n$ is the size of the vocabulary, 
$m$ is the dimension of the embedding vector. 
The embedding matrix $E=\left(\psi\left(w_1\right), \ldots, \psi\left(w_i\right)\right)^T$ is randomly initialized 
and learned together with the encoder during training. 
The translation's vector representation $v^s$ is obtained from the embedding matrix 
and inputted into the translation encoder to obtain the final embedding vector $e^s$. 
The hidden state $h_i^s$ for the $i$-th word in $s$ is calculated as follows: 
\begin{equation}
\label{equa6}
h_i^s={LSTM}\left(h_{i-1}^s, \boldsymbol{w}_i\right).
\end{equation}
In addition, 
they also employ an attention mechanism 
to alleviate the long dependency problem in long text sequences.  
The attention weight for each word is calculated as follows: 
\begin{equation}
\label{equa7}
\alpha_i^s=\frac{\exp \left(f\left(h_i^s\right) \cdot u^s\right)}{\sum_{j=1}^N \exp \left(f\left(h_j^s\right) \cdot u^s\right)},
\end{equation} 
where $f(\cdot)$ represents the linear layer; 
$u^s$ represents the context vector, 
which is the high-level representation of all words in $ s $. 
$\cdot$ represents the inner product of $h_i^s$ and $u^s$. 
$u^s$ is randomly initialized and jointly learned during training. 
The final embedding representation $e^s$ of $s$ is computed as follows:
\begin{equation}
\label{equa8}
e^s=\sum_{j=1}^{N^s} \alpha_i^s \cdot h_i^s.
\end{equation}

\subsubsection{Based on Graph Neural Networks}
As mentioned earlier, 
the code can be represented as a graph structure. 
This representation is superior to a sequence representation as it retains more information about the code. 
Some approaches use Graph Neural Networks (GNNs) to encode the graph structure, 
which is constructed based on structures such as AST and CFG \cite{zeng2023degraphcs, ling2021deep, liu2023graphsearchnet}.

Zeng et al. \cite{zeng2023degraphcs} construct a variable-based flow graph (VFG) 
based on the data dependencies and control dependencies of the source code. 
VFG is a directed graph with multiple types of edges, represented as $G=(V, E)$. 
They then use a GGNN model, which is well suited for handling graph-structured data, 
to learn the vector representation of code. 
Here, 
$V$ represents a set of nodes $\left(v, l_v\right)$, 
$E$ represents a set of edges $\left(v_i, v_j, l_{\left(v_i, v_j\right)}\right)$, 
and $l_v$ represents the label of node $v$, which consists of variables in the IR instruction. 
$l_{\left(v_i, v_j\right)}$ represents the label of the edge from $v_i$ to $v_j$, 
including data dependency and control dependency. 
GGNN learns the vector representation of $G$ through the message passing mechanism. 
In each iteration, 
each node $v_i$ receives a message $m_t^{v_j \mapsto v_j}=W_{l_{\left(v_i, v_j\right)}} h_{v_j}^{t-1}$ from its neighbor $v_j$, 
which is determined by the type of edge between them.
GGNN then aggregates all messages $m_t^i=\sum_{v_j \in {Neibour}\left(v_i\right)}\left(m_t^{v_j \mapsto v_i}\right)$ from neighbors of $v_i$, 
and updates the embedding $h_{v_i}^t=G R U\left(m_t^i, h_{v_i}^{t-1}\right)$ of each node $v_i$ using GRU. 
Considering that different nodes contribute differently to code semantics, 
they use an attention mechanism to calculate the importance of different nodes:
\begin{equation}
  \label{equa9}
  \alpha_i={sigmoid}\left(f\left(h_{v_i}\right) \cdot u_{v f g}\right),
\end{equation}
where $f(\cdot)$ is a linear layer and $u_{v f g}$ represents the context vector, 
which is a high-level representation of the entire nodes in the graph, 
learned together during training. 
The final embedding of the entire graph is expressed as: 
\begin{equation}
  \label{equa10}
  h_{v f g}=\sum_{v_i \in V}\left(\alpha_i h_{v_i}\right).
\end{equation}

Given that query text has sequence characteristics as a natural language, 
Zeng et al. \cite{zeng2023degraphcs} use LSTM to encode a query into a vector space and calculate its similarity with the code's semantic vector.  
Different from this, 
Ling et al. \cite{ling2021deep} design a unified graph structure for both  query and code.  
They then use RGCN to encode text graph and code graph, which is a variant of GNN. 
Given a code graph $G_e=\left(\mathcal{V}_e, \mathcal{E}_e, \mathcal{R}_e\right)$, 
to calculate the updated embedding vector $\mathbf{e}_i$ of each node $e_i$ in the code graph $G_e$, 
RGCN defines the propagation model as follows: 
\begin{equation}
  \label{equa11}
  \mathbf{e}_i^{(l+1)}={ReLU}\left(W_{\Theta}^{(l)} \mathbf{e}_i^{(l)}+\sum_{r \in \mathcal{R}_e} \sum_{j \in \mathcal{N}_i^r} \frac{1}{\left|\mathcal{N}_i^r\right|} W_r^{(l)} \mathbf{e}_j^{(l)}\right),
\end{equation} 
where $\mathbf{e}_i^{(l+1)}$ represents the updated embedding vector of node $\mathbf{e}_i$ in the $(l+1)$th layer of RGCN, 
$\mathcal{R}_e$ represents a set of relations (that is, the types of edge), 
$\mathcal{N}_i^r$ is a set of neighbors of node $e_i$ under the edge type $r \in \mathcal{R}_q$, 
and $W_{\Theta}^{(l)}$ and $W_r^{(l)}$ are the parameters that the RGCN model needs to learn. 
By encoding the graph structure of the code with the RGCN model, 
they obtain the node embedding $\mathbf{X}_e=\left\{\mathbf{e}_j\right\}_{j=1}^N \in \mathbb{R}^{(N, d)}$ of the code graph.  
The embedding $\mathbf{X}_q=\left\{\mathbf{q}_i\right\}_{i=1}^M \in \mathbb{R}^{(M, d)}$ of the query text graph can be obtained similarly.

\subsubsection{Based on Transformers}
In recent years, 
large pre-trained models have brought significant improvements to many NLP tasks. 
Many approaches train deep learning models on massive plain text data using self-supervised objectives, 
with the Transformer neural network architecture being the most prominent. 
Transformer contains multiple self-attention layers and can continuously learn in an end-to-end manner through gradient descent, 
because each of its components is differentiable. 
The success of pre-training models in NLP has inspired researchers to create code understanding pre-training models based on transformers, 
driving the growth of code intelligence. 

Feng et al. \cite{feng2020codebert} propose CodeBERT, 
the first large-scale natural language-programming language pre-training model for multiple programming languages. 
During pre-training, 
they use special tokens to splice natural language text sequences and code token sequences, 
which are fed into a multi-layer Transformer-based CodeBERT. 
The model learns the semantic relationship between natural language and programming language 
through Masked Language Modeling (MLM) and Replaced Token Detection (RTD) tasks, 
ultimately yielding a general vector for code understanding tasks.

Some methods aim to provide a more comprehensive understanding of code 
by feeding multiple modal representations of the source code into the Transformer.  
For instance, 
Guo et al. \cite{guo2020graphcodebert} combine the sequence of variables extracted from the data flow graph 
with the sequence of code tokens, 
and feed both into the multi-layer transformer-based GraphCodeBERT. 
They then use Masked Language Modeling (MLM), Edge Prediction (EP) and Node Alignment (NA) tasks 
to guide the model to learn the code structure and data dependencies. 
Additionally, Guo et al. \cite{guo2022unixcoder} merge the serialized AST with the comment text sequence 
and feed both into the multi-layer Transformer-based UnixCoder. 
They then use Masked Language Modeling (MLM), Unidirectional Language Modeling (ULM), DeNoiSing (DNS), Multi-modal Contrastive Learning (MCL), 
and Cross-Modal Generation (CMG) to learn the syntactic information of the code 
and enhance the understanding of the code semantics.
Wang et al. \cite{wang2022code} add not only the serialized AST, which reflects the grammatical information of the code, 
but also the CFG sequence, which reflects the logical information, 
and the token sequence of code fragments that have different implementation but the same semantics. 
All these are fed into the CODE-MVP. 
They then use Multi-View Contrastive Learning (MVCL), Fine-Grained Type Inference (FGTI), and Multi-View Masked Language Modeling (MMLM) tasks 
to help the model learn the structural information of the code. 

To ensure accurate code search results for each query, 
it's essential for the model to make the query and correct code vectors as close as possible in the shared vector space, 
and as far away as possible from incorrect code vectors. 
To achieve this, 
some researchers integrate contrastive learning into the Transformer network architecture 
and enhance the performance of code search engines by constructing positive and negative samples. 
For instance, 
Huang et al. \cite{huang2021cosqa} form negative sample pairs by randomly selecting the query and code within the same batch. 
Meanwhile, 
they generate positive sample pairs by duplicating the query, 
meaning they rephrase the query without altering its semantics. 
Developers often split a complete comment over several lines to improve readability. 
Inspired by this, 
Li et al. \cite{li2022coderetriever} combine consecutive comment lines into a single line
to make the most of code snippets without comments to form positive sample pairs for contrastive learning. 
When multiple modalities of code are available, 
Wang et al. \cite{wang2021syncobert} combine different modalities to create positive sample pairs 
and use both in-batch and cross-batch sampling methods to generate negative sample pairs.

Transformer is favored by researchers for its general and powerful modeling capabilities. 
To improve code semantic understanding, 
researchers have explored various pre-training tasks such as MLM, RTD, EP, and FGTI, etc. 
to guide model learning and 
enable the model to learn the grammatical and structural information of the code.

\subsubsection{Mixed Mode} 
The structures used to represent different modes of the code may vary, 
and some methods choose the proper model to deal with each mode accordingly. 
For instance, 
Wan et al. \cite{wan2019multi} extract AST and CFG as code representations, 
and propose Tree-LSTM for the tree structure of AST. 
Compared with the traditional LTSM unit, 
the Tree-LSTM unit contains multiple forgetting gates. 
For the directed graph structure of the CFG, 
they employ the Gated Graph Neural Network (GGNN) for encoding, 
and the Gated Recurrent Unit (GRU) for updating the hidden state of each vertex. 
Finally, they obtain the overall embedding vector of CFG by aggregating the embeddings of all vertices.

\subsection{Interaction Between Code and Query}
The existence of a semantic gap between natural language queries and code snippets has posed a significant challenge. 
However, several methods have emerged to bridge this gap by effectively modeling the interaction between the two, 
thus enhancing the understanding of their respective semantics \cite{haldar2020multi, xu2021two, arakelyan2022ns3, cheng2022csrs, hu2022cssam}. 
For instance, 
leveraging the widely adopted approach of calculating the overall similarity between queries and code, 
Haldar et al. \cite{haldar2020multi} additionally introduced the concept of evaluating local similarity between the two components. 
This perspective provides valuable insights into the finer-grained aspects of their correlation. 
Xu et al. \cite{xu2021two} proposed an innovative two-stage attention network architecture. 
In the initial stage, the semantics of both the query and the code are extracted. 
Subsequently, in the second stage, 
a joint attention mechanism is employed to facilitate the interaction between the two, 
enabling the capture of their semantic relevance. 
This approach presents a significant advancement in bridging the gap between natural language queries and code snippets. 
Similarly, 
Cheng and Kuang \cite{cheng2022csrs} combined neural information retrieval (IR) and semantic matching to enhance the interaction between queries and code. 
Their approach involved capturing two matching signals simultaneously. 
Firstly, neural IR captured keyword matching signals, 
encompassing words, terms, and phrases, within query-code pairs. 
Secondly, semantic matching employed a joint attention mechanism to simultaneously focus on description attention and code attention, 
thereby acquiring comprehensive semantic vector representations for both components. 
Particularly, 
Arakelyan et al. \cite{arakelyan2022ns3} proposed a meticulous approach wherein they parse the query by leveraging distinct part-of-speech roles to decompose it into concise semantic phrases. 
Specifically, nouns and noun phrases are associated with data entities within the code, 
while verbs depict operations or transformations performed on those entities. 
The interaction between the query and code is facilitated through an entity discovery module and an action module. 
The entity discovery module receives a string referencing a data entity and aims to identify code tokens that exhibit strong correlation with the given string. 
The output of this module is then employed as input for the action module, 
enabling prediction of the target entity object for the intended operational behavior. 
This comprehensive methodology offers valuable insights into enhancing the understanding and alignment between natural language queries and code snippets.

\subsection{Summary}
Source code can be represented in various modes, 
including code token sequences, Abstract Syntax Trees (AST), Control Flow Graphs (CFG), Data Flow Graphs (DFG), and Program Transformation (PT). 
Code token sequences are commonly utilized for extracting textual features of the code, 
while AST and CFG are frequently employed for extracting the structural features of the code. 
As Table \ref{table:overview} demonstrates, 
the multiple modes of code are mainly fed into the network in two forms of data structures: sequence and graph. 
The information from these different modes can complement each other, 
enabling the model to fully grasp the semantic information of the code. 
Based on the type of input data structure, 
an appropriate sequence embedding model or graph embedding model is selected to obtain the code vector representation.
As depicted in Figure 5, 
the predominant choice for code representation is source code token sequence (STS), 
while Transformer architecture has emerged as the widely adopted code encoder of choice.
Not all modalities will have equal impact on the final representation of the code. 
By aggregating embedding vectors from different modes of the source code, 
assigning attention weights, and taking the weighted sum, 
a more semantically rich code vector representation can be obtained. 
Furthermore, 
it is crucial to foster a fine-grained interaction between the query and code, 
facilitating a more robust learning of their intricate semantics.

\begin{table}[H]
  \caption{\textbf{Overview of approaches for code representation and code vectorization.}}
  \centering
  \scalebox{0.7}{
  \begin{tabular}{ccccccccccc}
  \toprule
  \multirow{2}{*}{Year} & \multirow{2}{*}{Work}            & \multicolumn{4}{c}{Code Representation} & \multicolumn{4}{c}{Code Vectorization} & \multirow{2}{*}{Interaction}\\
  \cmidrule(lr){3-6}\cmidrule(lr){7-10}
                                                   &  &    STS     &    MTS     &       FTS      & Tree/Graph & Word Embedding &    RNN     &    GNN     & Transformer  & \\
  \midrule
  2018 & CODEnn \cite{gu2018deep}                     &            &            &  \Checkmark    &            &                & \Checkmark &            &              & \\
  \midrule
  2018 & NCS \cite{sachdev2018retrieval}              &            &            &  \Checkmark    &            &  \Checkmark    &            &            &              & \\
  \midrule
  2019 & UNIF \cite{cambronero2019deep}               & \Checkmark &            &                &            &  \Checkmark    &            &            &              & \\
  \midrule
  2019 & MMAN \cite{wan2019multi}                     & \Checkmark &            &                & \Checkmark &                & \Checkmark & \Checkmark &              & \\
  \midrule
  2019 & CoaCor \cite{yao2019coacor}                  & \Checkmark &            &                &            &                & \Checkmark &            &              & \\
  \midrule
  2020 & CodeBERT \cite{feng2020codebert}             & \Checkmark &            &                &            &                &            &            &  \Checkmark  & \\
  \midrule
  2020 & AdaCS \cite{ling2020adaptive}                & \Checkmark &            &                &            &                & \Checkmark &            &              & \\
  \midrule
  2020 & MP-CAT \cite{haldar2020multi}                &            & \Checkmark &                &            &                & \Checkmark &            &              & \Checkmark \\
  \midrule
  2020 & ${TranS}^3$ \cite{wang2020trans}             &            &            &                & \Checkmark &                &            &            &  \Checkmark  & \\
  \midrule
  2020 & Zhao and Sun \cite{zhao2020adversarial}      & \Checkmark &            &                &            &                & \Checkmark &            &              & \\
  \midrule
  2020 & CO3 \cite{ye2020leveraging}                  & \Checkmark &            &                &            &                & \Checkmark &            &              & \\
  \midrule
  2021 & CRaDLe \cite{gu2021cradle}                   &            & \Checkmark &                &            &                & \Checkmark &            &              & \\
  \midrule
  2021 & GraphCodeBERT \cite{guo2020graphcodebert}    &            & \Checkmark &                & \Checkmark &                &            &            &  \Checkmark  & \\
  \midrule
  2021 & DGMS \cite{ling2021deep}                     &            &            &                & \Checkmark &                &            & \Checkmark &              & \\
  \midrule
  2021 & CoCLR \cite{huang2021cosqa}                  & \Checkmark &            &                &            &                &            &            &  \Checkmark  & \\
  \midrule
  2021 & DOBF \cite{lachaux2021dobf}                  & \Checkmark &            &                &            &                &            &            &  \Checkmark  & \\
  \midrule
  2021 & Gu et al. \cite{gu2021multimodal}            &            & \Checkmark &                &            &                &            &            &  \Checkmark  & \\
  \midrule
  2021 & TabCS \cite{xu2021two}                       &            & \Checkmark &                &            &                &            &            &  \Checkmark  & \Checkmark \\
  \midrule
  2021 & MuCoS \cite{du2021single}                    & \Checkmark &            &                &            &                &            &            &  \Checkmark  & \\
  \midrule
  2021 & SynCoBERT \cite{wang2021syncobert}           &            & \Checkmark &                &            &                &            &            &  \Checkmark  & \\
  \midrule
  2022 & Salza et al. \cite{salza2022effectiveness}   & \Checkmark &            &                &            &                &            &            &  \Checkmark  & \\
  \midrule
  2022 & G2SC \cite{shi2022better}                    &            & \Checkmark &                &            &                & \Checkmark &            &  \Checkmark  & \\
  \midrule
  2022 & CSRS \cite{cheng2022csrs}                    &            &            &   \Checkmark   &            &   \Checkmark   &            &            &              & \Checkmark \\
  \midrule
  2022 & TranCS \cite{sun2022code}                    & \Checkmark &            &                &            &                & \Checkmark &            &              & \\
  % \midrule
  % 2022 & cpt-code \cite{neelakantan2022text}          & \Checkmark &            &                &            &                &            &            &  \Checkmark  & \\
  \midrule
  2022 & Wang et al. \cite{wang2022bridging}          & \Checkmark &            &                &            &                &            &            &  \Checkmark  & \\
  \midrule
  2022 & ${NS}^3$ \cite{arakelyan2022ns3}             & \Checkmark &            &                &            &                &            &            &  \Checkmark  & \Checkmark \\
  \midrule
  2022 & CodeRetriever \cite{li2022coderetriever}     & \Checkmark &            &                &            &                &            &            &  \Checkmark  & \\
  \midrule
  2022 & CDCS \cite{chai2022cross}                    & \Checkmark &            &                &            &                &            &            &  \Checkmark  & \\
  \midrule
  2022 & CSSAM \cite{hu2022cssam}                     & \Checkmark &            &                & \Checkmark &                & \Checkmark & \Checkmark &              & \Checkmark \\
  \midrule
  2022 & SCodeR \cite{li2022soft}                     & \Checkmark &            &                &            &                &            &            &  \Checkmark  & \\
  \midrule
  2022 & SPT-Code \cite{niu2022spt}                   &            & \Checkmark &                &            &                &            &            &  \Checkmark  & \\
  \midrule
  2022 & Li et al. \cite{li2022exploring}             & \Checkmark &            &                &            &                &            &            &  \Checkmark  & \\
  \midrule
  2022 & CTBERT \cite{han2022towards}                 &            & \Checkmark &                &            &                &            &            &  \Checkmark  & \\
  \midrule
  2022 & CODE-MVP \cite{wang2022code}                 &            & \Checkmark &                &            &                &            &            &  \Checkmark  & \\
  \midrule
  2022 & UniXcoder \cite{guo2022unixcoder}            &            & \Checkmark &                &            &                &            &            &  \Checkmark  & \\
  \midrule
  2023 & deGraphCS \cite{zeng2023degraphcs}           &            &            &                & \Checkmark &                &            & \Checkmark &              & \\
  \midrule
  2023 & MulCS \cite{ma2023mulcs}                     &            &            &                & \Checkmark &                &            & \Checkmark &              & \\
  \midrule
  2023 & GraphSearchNet \cite{liu2023graphsearchnet}  &            &            &                & \Checkmark &                &            & \Checkmark &              & \\
  \bottomrule
  \end{tabular}
  }
  \label{table:overview}
\end{table} 

\begin{figure}[htbp]
	\centering
	\begin{minipage}[c]{0.48\textwidth}
		\centering
		\includegraphics[width=\textwidth]{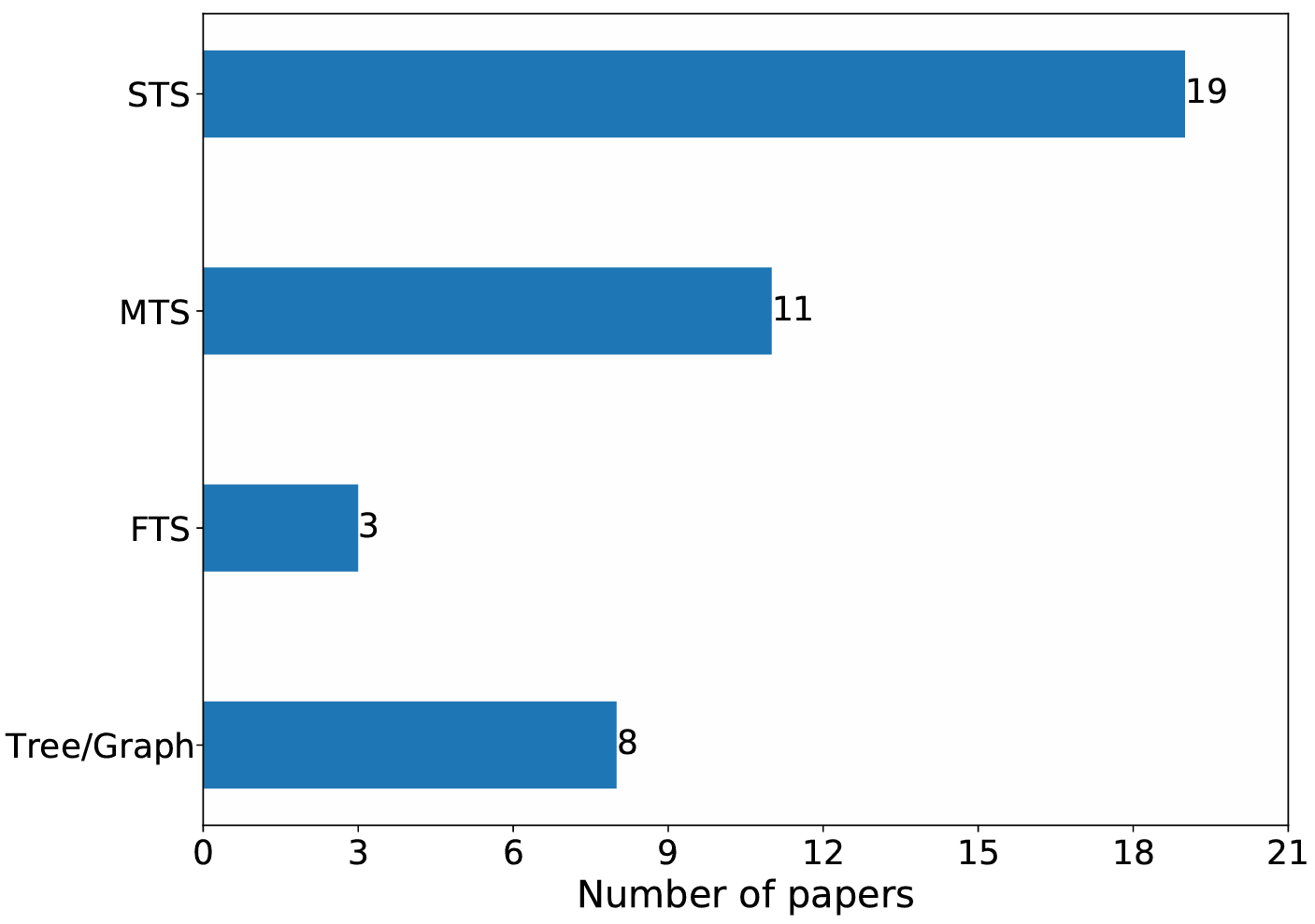}
		\subcaption{Code representation}
		\label{fig_E1_1}
	\end{minipage} 
	\begin{minipage}[c]{0.48\textwidth}
		\centering
		\includegraphics[width=\textwidth]{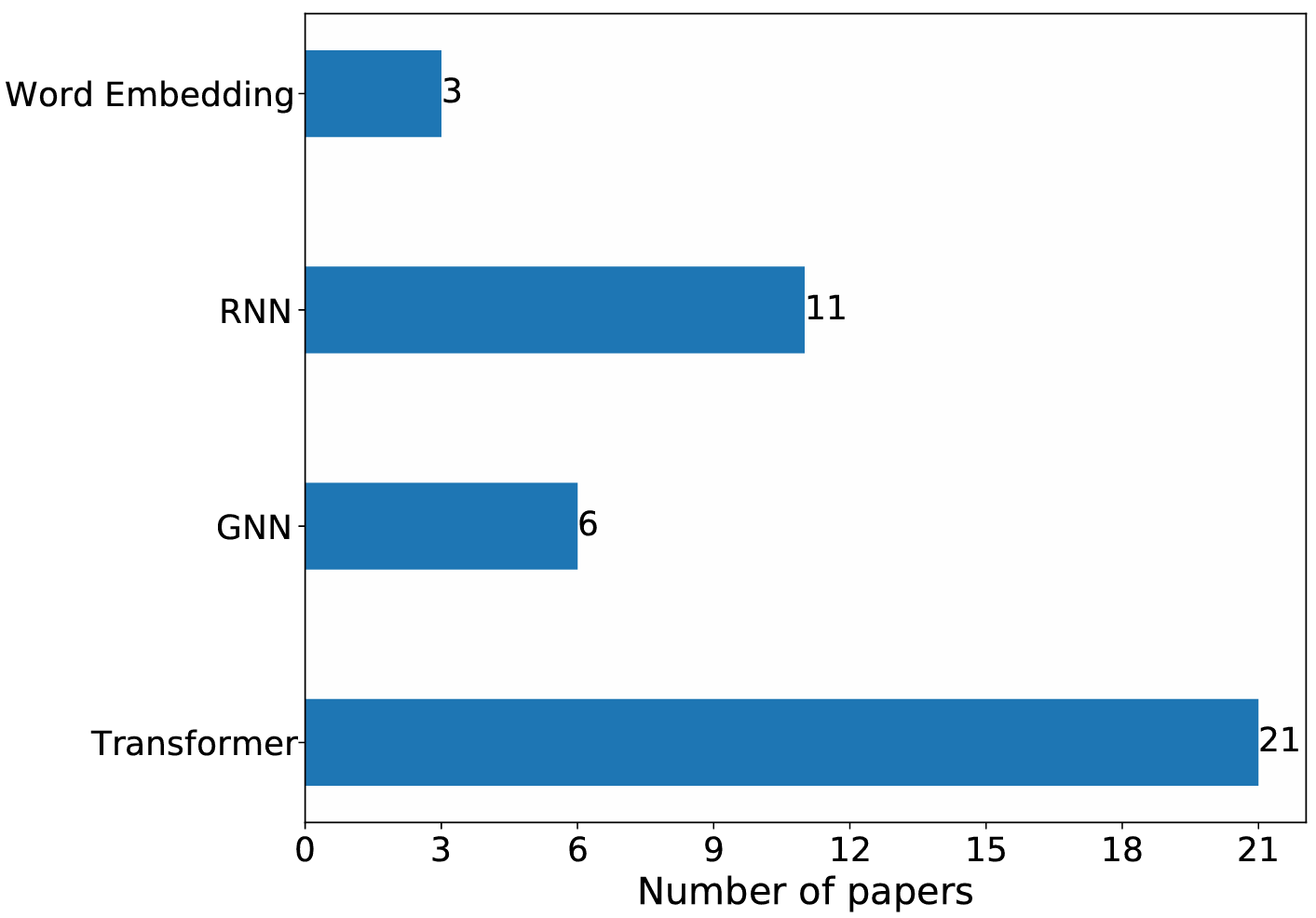}
		\subcaption{Code vectorization}
		\label{fig_E1_2}
	\end{minipage}
	\caption{Code representation and code vectorization.}
	\label{fig_E1}
\end{figure}

\newtcolorbox{RQ2box}{colback=white, colframe = black}
\begin{RQ2box}
  \textbf{\textit{Summary of answers to RQ2: }}
  \begin{itemize}
    \item \textit{An array of diverse modalities, 
including Abstract Syntax Trees (AST), Data Flow Graphs (DFG), Control Flow Graphs (CFG), Program Trees (PT), and other code representations, 
can be leveraged to effectively augment the model's ability to learn the intricate semantics of the code.}
    \item \textit{When feeding the code into the encoder, 
it is common practice to represent it as a sequence or a graph. 
Among these forms, the source code token sequence (STS) is the prevailing choice in most scenarios.}
    \item \textit{Transformer is the most popular code encoder in the past 6 years.}
    \item \textit{Facilitating a fine-grained interaction between the query and code can enhance the model's ability to grasp their semantics.}
  \end{itemize}
\end{RQ2box}

\section{Training Method (RQ3)}
As previously stated, 
the core problem of code search based on deep learning is 
to employ a model $f_\theta(q, c)=E_{\theta_1}(q) \cdot E_{\theta_2}(c)$ to estimate the relevance scores of query and code, 
and then sort according to the scores to obtain the code that matches the query. 
Sections 3 and 4 elaborate on the utilization of deep learning models to obtain representations for both the query and code. 
This section provides an overview of the existing techniques for training the model parameter $\theta$. 
The section is organized as follows: 
first, 
the pre-training technology of the Transformer-based large model relevant to code intelligence is introduced, 
encompassing three categories: sequence tasks, code structure tasks, and multimodal tasks. 
Second, 
the training of the code search model is described. 
Finally, 
alternative training methods for other scenarios, such as meta-learning and data enhancement, are introduced.

\subsection{Pre-training of Code Intelligence Large Models}
Encoding query and code using Transformer encoder is 
an important encoding strategy in code search tasks. 
Pre-trained Transformers have proven their effectiveness in many fields \cite{devlin2018bert, liu2019roberta}. 
Large-scale pre-training enables the model to start from a better initial state, 
thereby facilitating efficient training and fine-tuning for downstream tasks.  
At present, 
there have been many works to introduce pre-training technology 
into the field of code intelligence, 
resulting in the design of various pre-training tasks. 
These tasks can be classified into three categories: 
sequence tasks, code structure tasks, and multi-modal matching tasks. 
Table \ref{table:pre-training} summarizes the approaches along these three dimensions.
Sequence tasks treat the input as a sequence 
and train by using popular NLP tasks. 
Code structure tasks build upon sequences and utilize the relationships between tokens in a code structure graph for training. 
Multi-modal matching tasks try to 
introduce contrastive learning into pre-training, 
and use different modalities to construct matching positive samples. 

\begin{table}[H]
  \caption{\textbf{Overview of pre-training tasks.}}
  \centering
  \resizebox{\textwidth}{!}{
  \begin{tabular}{cccccccccccc}
  \toprule
  \multirow{2}{*}{Year} & \multirow{2}{*}{Work}            & \multicolumn{4}{c}{Sequence Tasks} & \multicolumn{3}{c}{Code Structure Tasks} & \multicolumn{3}{c}{Multimodal Matching Tasks} \\
  \cmidrule(lr){3-6}\cmidrule(lr){7-9}\cmidrule(lr){10-12}
                                                &  &    MLM     &    RTD     &       ULM       &   DNS      &     EP     &    IP      &    ICP     &   MMP      &  MCCL       &  CMG       \\
  \midrule
  2020 & CodeBERT \cite{feng2020codebert}          & \Checkmark  & \Checkmark &                &            &            &            &            &            &             &            \\
  \midrule
  2021 & GraphCodeBERT \cite{guo2020graphcodebert} &  \Checkmark &            &                &            & \Checkmark &            &            &            &             &             \\
  \midrule
  2021 & Syncobert \cite{wang2021syncobert}        &  \Checkmark &            &                &            & \Checkmark & \Checkmark &            &            &             &              \\
  \midrule
  2022 & CODE-MVP \cite{wang2022code}              &  \Checkmark &            &                &            &            & \Checkmark &            &            &             &              \\
  \midrule
  2022 & CTBERT \cite{han2022towards}              &             &            &                &            & \Checkmark &            &            &            &             &              \\
  \midrule
  2022 & SPT-Code \cite{niu2022spt}                &             &            &                & \Checkmark &            &            & \Checkmark & \Checkmark &             &             \\
  \midrule
  2022 & DOBF \cite{lachaux2021dobf}               &             &            &                &            &            &            & \Checkmark &            &             &             \\
  \midrule
  2022 & SCodeR \cite{li2022soft}                  &             &            &                &            &            &            &            &            & \Checkmark  &             \\
  \midrule
  2022 & UniXcoder \cite{guo2022unixcoder}         &  \Checkmark &            &  \Checkmark    & \Checkmark &            &            &            &            &             & \Checkmark \\
  \bottomrule
  \end{tabular}
  }
  \label{table:pre-training}
\end{table}

\subsubsection{Sequence Tasks} 
Sequences are the most straightforward and efficient way of representing code and text. 
Data structures like AST and CFG can be transformed into sequences through a serialization algorithm. 
There are many established pre-training tasks in NLP for encoding sequences in Transformer models, 
which can be directly introduced into code intelligence.

\textbf{Masked Language Model (MLM)} \cite{feng2020codebert, guo2020graphcodebert, wang2021syncobert, guo2022unixcoder, wang2022code}. 
Given a token sequence 
$X=\left\{[cls], x_1, x_2, \cdots \right.$
$\left. , x_N,[end]\right\}$, 
some tokens are randomly masked as $[mask]$  
and a new token sequence is denoted as 
$X^{mask }=\left\{x_1, x_2, \cdots,[mask], \cdots, x_N\right\}$. 
$X^{mask }$ is input ted into the model, 
and the objective-task can restore the original tokens 
through the representation at the output layer. 
The loss function can be expressed as follows: 
\begin{equation}
  \label{equa12}
  {Loss}_{MLM}=-\sum_i \log p\left(x_i \mid X^{mask }\right).
\end{equation}

MLM can be extended to the input of multiple modal splicing, 
that is, 
splicing query, code, and other modalities of code (such as flattened AST and CFG sequences). 
This results in a new sequence 
$X=\left\{[cls], x_1^{(1)}, x_2^{(1)}, \cdots, x_N^{(1)},[{sep}], x_1^{(2)}, x_2^{(2)}, \cdots, x_M^{(2)},[end]\right\}$, 
on which the MLM task is performed. 
In some literatures, 
this type of Masked Language Model that combines multiple modalities is referred to as Multi-Modal Masked Language Modeling (MMLM) \cite{wang2021syncobert}.

\textbf{Replaced Token Detection (RTD)} \cite{feng2020codebert}. 
A token sequence is given, 
and some of its tokens are randomly replaced with others, creating a new token sequence, 
denoted as  
$X^{corrupt }=\left\{[cls], x_1^{\prime}, x_2^{\prime}, \cdots, \right.$
$\left. x_N^{\prime},[end]\right\}$. 
The task makes predictions at the output layer regarding whether a token in $X^{corrupt }$ is a replacement token.
The loss function can be expressed as follows: 
\begin{equation}
  \label{equa13}
  {Loss}_{RTD}=-\sum_i\left(\delta(i) \log \left(i \mid X^{corrupt }\right)+(1-\delta(i))(1-
\left.\log p\left(i \mid X^{corrupt }\right)\right)\right).
\end{equation}
When the token is not replaced (that is, $x_i^{\prime}=x_i$), $\delta(i)=1$, otherwise $\delta(i)=0$. 

\textbf{Unidirectional Language Modeling (ULM)} \cite{guo2022unixcoder}. 
ULM predicts the tokens in the sequence from start to end 
and is a widely used auxiliary task for training the Transformer decoder. 
In some models that train both the encoder and decoder \cite{guo2022unixcoder}, 
ULM is selected to train the decoder. 
The loss function can be expressed as follows: 
\begin{equation}
  \label{equa14}
  {Loss}_{ULM}=-\sum_i \log p\left(x_i \mid X[<i]\right),
\end{equation}
where $X[<i]$ represents the tokens before the $i$-th token.

\textbf{DeNoiSing (DNS)} \cite{guo2022unixcoder, niu2022spt}. 
The DeNoiSing pre-training objective uses $[mask_i]$ to mask multiple spans 
in the input tokens sequence to obtain $X^{mask}$, 
where $i$ represents the $i$-th masked span. 
All $X^{mask}$ are concatenated and represented as $Y=\left\{y_1, y_2, \cdots\right\}$.  
The corresponding output can be reconstructed 
through the encoder-decoder framework, 
and the objective function can be expressed as follows: 
\begin{equation}
  \label{equa15}
  {Loss}_{DNS}=-\sum_i \log \left(y_i \mid X^{mask }, Y[<i]\right).
\end{equation}

\subsubsection{Code Structure Tasks} 
The structured representation of code, such as AST and CFG, often holds a lot of information, 
such as the variable jump relationships, code execution order, etc. 
These information can be mined to design auxiliary tasks that aid in pre-training code intelligence models.

\textbf{Edge Prediction (EP)} \cite{guo2020graphcodebert, wang2021syncobert, han2022towards}. 
When encoding structured code representations using a Transformer, 
it's common to flatten it into a sequence 
$X=\left\{[cls], x_1, x_2, \cdots, x_N,[end]\right\}$. 
After that, 
the connection relationship of the nodes in the original graph/tree 
can be inherited into the tokens sequence. 
For example, if node $i$ points to node $j$ in the original graph, 
then token $x_i$ and token $x_j$ will inherit this connection, allowing it to guide the model's pre-training.  
The probability $p\left(e_{i j} \mid X\right)$ of 
whether there is an edge between two tokens is measured by 
the similarity of representations $\vec{e}_{i}$ and $\vec{e}_{j}$ of $x_i$ and $x_j$, namely $<\vec{e}_i, \vec{e}_j>$. 
The objective function of EP can be expressed as follows: 
\begin{equation}
  \label{equa16}
  {Loss}_{EP}=-\sum_{i j}\left(\delta\left(e_{ij}\right) \log \left(e_{i j} \mid X\right)+\left(1-\delta\left(e_{ij}\right)\right)(1-\left.{logp}\left(e_{i j} \mid X\right)\right)\right).
\end{equation}
When there is an edge between token $x_i$ and token $x_j$, 
$\delta\left(e_{i j}\right)=1$, otherwise $\delta\left(e_{i j}\right)=0$. 
Besides single-modal structures, 
cross-modal structures can also be used to construct edges between tokens,  
for example, 
by combining code sequence tokens with data flow tokens.
There is a correspondence between them, 
and edges can be constructed according to the corresponding relationship.

\textbf{Identifier Prediction (IP)} \cite{wang2022code, wang2021syncobert}. 
The tokens in the code snippet can be classified into two types: 
identifiers (such as variable and function names, etc.) 
and non-identifiers (keywords that indicate grammatical operations). 
This classification can be performed through AST-based code analysis. 
The information helps construct auxiliary tasks to 
determine whether a token is an identifier. 
The probability $p(i\mid X)$ that a token is an identifier can be predicted 
based on its representation through a simple transformation, 
namely $\sigma\left(\vec{w}^{T} \cdot \vec{e}_{i}\right)$, 
where $\sigma$ is a sigmoid function that maps the score to 0-1. 
The objective function of IP can be expressed as follows: 
\begin{equation}
  \label{equa17}
  {Loss}_{IP}=-\sum_i(\delta(i) \log p(i \mid X)+(1-\delta(i))(1-\log p(i \mid X))).
\end{equation}
If the node is an identifier, $\delta(i)=1$; otherwise $\delta(i)=0$.

\textbf{Identifier Content Prediction (ICP)} \cite{lachaux2021dobf, niu2022spt}. 
Identifiers such as function names and variable names carry significant information that can be considered as a general expression of functions. 
To allow the model to learn this high-level semantic information, 
the following proxy tasks can be designed:
masking the identifier as $[mask]$, and predicting the masked content. 
Compared to randomly masking tokens, 
accurately masking identifiers by analyzing the code structure 
increases the difficulty of the task and makes the model learn higher-level information. 
When masking identifiers, 
to avoid the model dependence on the same variable name appearing in the context for inference, 
all the same tokens can be replaced by $[mask_i]$ at the same time, 
where $i$ represents the $i$-th replaced identifier. 
When actually designing the training task, 
$[mask_i]$ can be connected to the masked token. 
All masked information is concatenated as output  
and then the model is trained using the encoder-decoder framework. 
The corresponding objective function can be expressed as follows: 
\begin{equation}
  \label{equa18}
  {Loss}_{ICP}=-\sum_i \log p\left(y_i \mid X^{mask }, y[<i]\right).
\end{equation}

\subsubsection{Multimodal Matching Tasks}
From a multi-modal viewpoint, 
code intelligence is a multi-modal task involving natural language, 
code sequence representation, and code structure representation. 
At present, 
many studies are also examining this problem from a multi-modal perspective 
and designing pre-training tasks for multi-modal matching.

\textbf{Multi-Modal Matching Prediction (MMP)} \cite{niu2022spt}. 
The multiple modalities of code can be seen as 
different ways of describing information with the same semantics. 
MMP splices the descriptions of these different modalities to obtain $M^{(1)} / / M^{(2)} / / \cdots / / M^{(N)}$, 
where $M^{(i)}$ represents the token sequence composed of modality $i$. 
The number of modalities can be greater than or equal to 2, 
but attention should be paid to controlling the total amount, 
as sequences that are too long can result in a high computational complexity for the Transformer. 
Usually, $N$ is set to 2 or 3. 
Assuming $N=2$, 
the spliced sequence is denoted as $M_i^{(1)} / / M_j^{(2)}$, 
where $M_i^{(j)}$ represents the $j$-th modal sequence representation of code $i$. 
The training objective is to determine if the two modalities in the spliced sequence describe the same piece of code information. 
If they do, 
the label is $\delta(i j)=1$, otherwise it is $\delta(i j)=0$. 
The spliced sequence can be input ted to the model, 
and the probability of $i=j$ can be estimated by the representation of $[cls]$. 
The loss function can be expressed as follows:
\begin{equation}
  \label{equa19}
  {Loss}_{MMP}=-\sum_{i j}\left(\delta(i j) \log p\left(c l s \mid M_i^{(1)} / / M_j^{(2)}\right)+(1-\delta(i j))\left(1-\log p\left(c l s \mid M_i^{(1)} / / M_j^{(2)}\right)\right)\right).
\end{equation}

\textbf{Multi-modal Concatenation Contrastive Learning (MCCL)} \cite{li2022soft}. 
In multi-modal contrastive learning, 
the model encodes information from two modalities and calculates similarity in the representation space, namely 
$s_{i j}=<E\left(M_i^{(1)}\right), E\left(M_j^{(2)}\right)>$. 
However, this approach doesn't model the fine-grained interaction of $M_i^{(i)}$ and $M_i^{(j)}$ at the token level. 
To address this limitation, 
the token sequences of the two modalities can be spliced and encoded together with a single encoder, 
namely $s_{i j}=E\left(M_i^{(1)} / / M_j^{(2)}\right)$. 
Based on the similarity, the loss function is constructed as follows:
\begin{equation}
  \label{equa20}
  {Loss}_{MCCL}=-\sum_i \log \frac{\exp \left(s_{i i} / \tau\right)}{\sum_j \exp \left(s_{i j} / \tau\right)}.
\end{equation}
After training with this objective, 
researchers can continue to use $s_{i j}$ to weight the samples effectively in multi-modal contrastive learning,
thereby enhancing its performance.

\textbf{Cross Modal Generation (CMG)} \cite{guo2022unixcoder}. 
CMG aims to predict the information of another modal sequence through the information of one modal sequence. 
For example, the comment of the code can be predicted based on the function body sequence. 
If the source modality is denoted as $X$ and the target modality is denoted as $Y$, 
the objective function can be expressed as follows: 
\begin{equation}
  \label{equa21}
  {Loss}_{CMG}=-\sum_i \log p\left(y_i \mid X, Y[<i]\right).
\end{equation} 
CMG can be regarded as a generation task that utilizes multi-modal matching information, 
similar to an implicit matching task.

\subsubsection{Summary}
This section introduces ten pre-training tasks for code intelligence,  
covering code sequence, code structure, and code multimodality.
These tasks allow for training large models using extensive amounts of unlabeled data,  
leading to pre-trained models that can excel in various downstream tasks, such as code search, 
with minimal fine-tuning efforts. 
In practical applications, 
simple training tasks like MLM are often highly effective and play a fundamental role. 
Their simplicity and effectiveness make them a staple in most pre-training models, such as CodeBERT \cite{feng2020codebert} and GraphCodeBERT \cite{guo2020graphcodebert}.  
Complex tasks show limited improvement in the presence of simple training tasks as seen in some ablation experiments \cite{guo2022unixcoder},  
and they do not generalize well to all data and tasks \cite{wang2021syncobert}. 
Despite this, 
the variety of tasks also expands the possibilities for pre-training code intelligence, increasing its potential. 
Furthermore, 
for code search, pre-training with multimodal contrastive learning has produced better results due to its greater alignment with the downstream task \cite{li2022coderetriever}. 
This confirms that pre-training tasks perform better when they are consistent with the downstream tasks \cite{zhang2020pegasus}.

\subsection{Code Search Model Training/Fine-tuning}
The pre-training tasks for training large models for code understanding are described above.
Pre-training helps the parameters of the model reach a better state, 
which benefits the training of downstream tasks. 
This process is commonly referred to as fine-tuning. 
It is important to note that pre-training goals may not align with the target task, 
but training/fine-tuning should be focused on the target task and evaluated using the target task's evaluation method. 
Models that do not require pre-training, like Graph Neural Networks, also require design goals for optimization. 
For simplicity and consistency with models without pre-training, we refer to fine-tuning as training below. 

During training, 
the model is guided by task-specific labels, known as supervision information. 
For code search, supervision information consists of paired (query, code) samples. 
Based on different training scenarios, 
we categorize existing code search training into discriminative model training and generative model training. 
Table \ref{table:fine-tuning} summarizes the approaches along these two dimensions. 
The discriminative model is the most commonly used in code search and models the similarity $s_{qc}=f_\theta(q, c)$. 
The generative model in code search mainly involves a Variational Auto-Encoder, 
encoding the input query/code into a distribution, 
and then decoding the distribution back to the original query/code. 
By designing a suitable training strategy for the encoder-decoder framework, 
the trained Variational Auto-Encoder model has a good generalization ability for its distribution in the latent space, 
and similarity can be calculated based on the matching of the mean vector in the latent space distribution. 

\subsubsection{Discriminative Model Training} 
There are three types: point-level, pair-level, and sequence-level, which will be introduced separately.

\textbf{Point-wise} \cite{huang2021cosqa, arakelyan2022ns3}. 
In this task, 
paired (query, code) sample pairs are labeled as 1, 
while unpaired (query, code) sample pairs are labeled as 0. 
Therefore, 
the training data can be viewed as a set of (query, code, label) triplets, with labels being either 1 or 0. 
This task is a binary classification problem, 
and the model can be optimized using a binary classification objective such as Mean Squared Error loss:
\begin{equation}
  \label{equa22}
  {Loss}_{MSE}=\frac{1}{|D|} \sum_{(q, c) \in D}\left|\hat{y}_{qc}-y_{qc}\right|^2,
\end{equation}
where $y_{qc}$ represents the true label of the sample pair $(q,c)$, 
$\hat{y}_{qc}=\sigma\left(f_\theta(q, c)\right)$ represents the similarity score output by the model, 
$\sigma$ is the sigmoid function to ensure that the value of $\hat{y}_{qc}$ is between 0 and 1,  
and $D$ represents the sample set. 
During optimization, 
to maintain a balance between positive and negative samples, it is necessary to sample an equal number of negative samples. 

\begin{table}[H]
  \caption{\textbf{Overview of code search model training/fine-tuning.}}
  \centering
  % \resizebox{\textwidth}{!}{
  \scalebox{0.8}{
  \begin{tabular}{ccccccc}
  \toprule
  \multirow{2}{*}{Year} & \multirow{2}{*}{Work}            & \multicolumn{3}{c}{Discriminative Model Training}       & \multirow{2}{*}{Generative Model Training} & \multirow{2}{*}{Other} \\
  \cmidrule(lr){3-5}
                                                      &    &    Point-wise     &    Pair-wise     &      List-wise   &                                            &                     \\
  \midrule
  2018 & CODEnn \cite{gu2018deep}                          &                   & \Checkmark       &                  &                                            &                      \\
  \midrule
  2018 & Chen and Zhou \cite{chen2018neural}               &                   &                  &                  &           \Checkmark                       &                      \\
  \midrule
  2019 & MMAN \cite{wan2019multi}                          &                   & \Checkmark       &                  &                                            &                     \\
  \midrule
  2019 & CoaCor \cite{yao2019coacor}                       &                   & \Checkmark       &                  &                                            &                     \\
  \midrule
  2020 & AdaCS \cite{ling2020adaptive}                     &                   & \Checkmark       &                  &                                            &                      \\
  \midrule
  2020 & MP-CAT \cite{haldar2020multi}                     &                   & \Checkmark       &                  &                                            &                      \\
  \midrule
  2020 & ${TranS}^3$ \cite{wang2020trans}                  &                   &                  &  \Checkmark      &                                            &                      \\
  \midrule
  2020 & Zhao and Sun \cite{zhao2020adversarial}           &                   & \Checkmark       &                  &                                            &                      \\
  \midrule
  2020 & CO3 \cite{ye2020leveraging}                       &                   &                  &   \Checkmark     &                                            &                      \\
  \midrule
  2021 & DGMS \cite{ling2021deep}                          &                   & \Checkmark       &                  &                                            &                      \\
  \midrule
  2021 & CoCLR \cite{huang2021cosqa}                       &    \Checkmark     &                  &                  &                                            &                      \\
  \midrule
  2021 & TabCS \cite{xu2021two}                            &                   & \Checkmark       &                  &                                            &                      \\
  \midrule
  2021 & CRaDLe \cite{gu2021cradle}                        &                   & \Checkmark       &                  &                                            &                      \\
  \midrule
  2021 & Corder \cite{bui2021self}                         &                   & \Checkmark       &                  &                                            &                      \\
  \midrule
  2022 & TranCS \cite{sun2022code}                         &                   & \Checkmark       &                  &                                            &                     \\
  \midrule
  2022 & CSRS \cite{cheng2022csrs}                         &                   &                  &   \Checkmark     &                                            &                     \\
  \midrule
  2022 & G2SC \cite{shi2022better}                         &                   &  \Checkmark      &                  &                                            &                      \\
  \midrule
  2022 & Li et al. \cite{li2022exploring}                  &                   &                  &   \Checkmark     &                                            &                      \\
  % \midrule
  % 2022 & cpt-code \cite{neelakantan2022text}               &                   &                  &   \Checkmark     &                                            &                      \\
  \midrule
  2022 & CodeRetriever \cite{li2022coderetriever}          &                   &                  &   \Checkmark     &                                            &                     \\
  \midrule
  2022 & CDCS \cite{chai2022cross}                         &                   &                  &                  &                                            &     \Checkmark        \\
  \midrule
  2022 & ${NS}^3$ \cite{arakelyan2022ns3}                  &   \Checkmark      &                  &                  &                                            &                      \\
  \midrule
  2022 & Wang et al. \cite{wang2022bridging}               &                   &                  &                  &                                            &     \Checkmark        \\
  \midrule
  2023 & deGraphCS \cite{zeng2023degraphcs}                &                   & \Checkmark       &                  &                                            &                      \\
  \midrule
  2023 & KeyDAC \cite{park2023contrastive}                 &                   &                  &   \Checkmark     &                                            &                      \\
  \midrule
  2023 & CSSAM \cite{hu2022cssam}                          &                   & \Checkmark       &                  &                                            &                      \\
  \midrule
  2023 & MulCS \cite{ma2023mulcs}                          &                   & \Checkmark       &                  &                                            &                      \\
  \midrule
  2023 & GraphSearchNet \cite{liu2023graphsearchnet}       &                   &                  &   \Checkmark     &                                            &                      \\
  \midrule
  2023 & TOSS \cite{hu2023revisiting}                      &                   &                  &   \Checkmark     &                                            &                      \\
  \bottomrule
  \end{tabular}
  }
  \label{table:fine-tuning}
\end{table}

\textbf{Pair-wise} \cite{gu2018deep, yao2019coacor, wan2019multi, haldar2020multi, ling2020adaptive, zhao2020adversarial, ling2021deep, xu2021two, bui2021self, gu2021cradle, sun2022code, hu2022cssam, ma2023mulcs, zeng2023degraphcs}. 
For the paired (query, code) positive sample pair, 
the negative sample code can be randomly selected to 
construct a triplet $\left(q, c^{+}, c^{-}\right)$ 
called (query, positive sample code, negative sample code). 
The objective of training is to maximize the gap between 
positive sample $\left(q, c^{+}\right)$  and negative sample $\left(q, c^{-}\right)$. 
The loss function commonly used for this purpose is the Hinge loss, which can be represented as follows:
\begin{equation}
  \label{equa23}
  Loss_{hinge}=\sum_{\left(q, c^{+}, c^{-}\right) \in D} \max \left(0, \epsilon-s_{{qc}^{+}}+s_{{qc}^{-}}\right),
\end{equation}
where $s_{q c}=f_\theta(q, c)$ represents the similarity between query $q$ and code $c$, 
$\epsilon$ is a hyperparameter, and $D$ represents the sample set. 
For paired $(q,c)$ samples, negative samples $c^-$ are randomly selected to construct triplet sample $\left(q, c^{+}, c^{-}\right)$.

\textbf{List-wise} \cite{ye2020leveraging, wang2020trans, cheng2022csrs, li2022coderetriever, li2022exploring, liu2023graphsearchnet, hu2023revisiting}. 
For paired (query, code) positive sample pairs, 
several negative sample codes can be randomly selected to construct the sequence $\left(q, c^{+}, c_1^{-}, \cdots, c_{n-1}^{-}\right)$. 
The training goal is to optimize the similarity ranking of 
positive sample $c^+$ in the entire sequence $\left(c^{+}, c_1^{-}, \cdots, c_{n-1}^{-}\right)$ for the query $q$. 
The InfoNCE loss function can be utilized as the related loss function, 
viewing the problem as an $n$ classification task with the number of categories equal to the number of positive sample categories. 
After passing the similarity through softmax, 
the cross-entropy is used to construct the loss. 
The loss function can be expressed as: 
\begin{equation}
  \label{equa24}
  {Loss}_{InfoNCE}=-\sum_{\left(q, c^{+}, c_1^{-}, \cdots, c_{n-1}^{-}\right) \in D} \log \left(\frac{\exp \left(s_{q c^{+}} / \tau\right)}{\exp \left(s_{q c^{+}} / \tau\right)+\sum_j \exp \left(s_{q c_j^{-}} / \tau\right)}\right),
\end{equation}
where $\tau$ is the temperature hyperparameter. 
In practice, 
the batch negative sampling strategy is commonly used to generate negative samples, 
meaning that for a batch of positive sample pairs, other codes within the batch are considered as negative samples.  
For instance, 
\cite{li2022coderetriever} employs various strategies to sample negative samples to improve model performance, 
while \cite{li2022exploring} improves the representation space data and generates more positive and negative samples.

The above three optimization objectives are all designed for discriminative models in supervised learning. 
Among them, the sequence-level optimization objective is currently the most popular due to its effectiveness.

\subsubsection{Generative Model Training} 
The Variational Auto-Encoder (VAE) map the input $\vec{x}$ to a distribution in the hidden space through the function $f_\theta(\cdot)$ 
(assumed to be a Gaussian distribution, which can be represented by a mean and variance vector), 
and then the vector obtained from sampling the distribution is then mapped to the input space by the function $g_\phi(\cdot)$, hoping to reconstruct $\vec{x}$. 
A regularization term is added to make the latent space distribution closer to a standard Gaussian distribution.
The loss function of VAE is: 
\begin{equation}
  \label{equa25}
  {Loss}_{VAE}=\mathcal{R}(\vec{x} ; \theta, \phi)+KL\left(q_\theta(\vec{z} \mid \vec{x}), \mathcal{N}(0,1)\right),
\end{equation}
where $\mathcal{R}(\vec{x} ; \theta, \phi)$ denotes the error loss for reconstructing $\vec{x}$ 
through the encoder-decoder framework, 
and $KL\left(q_\theta(\vec{z} \mid \vec{x}), \mathcal{N}(0,1)\right)$ denotes 
the KL divergence of $q_\theta(\vec{z} \mid \vec{x})$ over $\mathcal{N}(0,1)$.

\begin{figure}[htbp]
  \centering
  \includegraphics[width=0.5\linewidth]{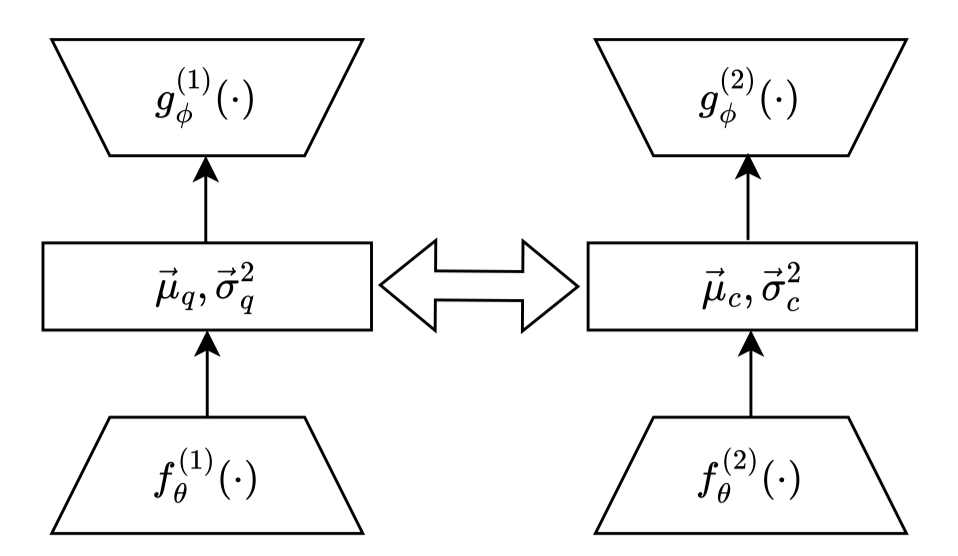}
  \caption{Bimodal Variational Auto-Encoder.}
  \label{VAE}
\end{figure}

The bimodal VAE, unlike general VAE, 
addresses the two modalities in code search and focuses on learning the matching relationships. 
The bimodal VAE has two improvements over the general VAE to make it more suitable for code search tasks \cite{chen2018neural}, 
as shown in Figure \ref{VAE}. 
(a) Bimodal encoder-decoder framework: 
this framework inputs both the query and the code into the respective encoder-decoder models (the parameters of each model can be the same), and they operate in parallel.  
(b) Cross-modal regularization term: 
the regularization involves minimizing the difference in distribution between two modes of a sample pair 
instead of using the standard normal distribution regularization term. 
This unifies the query and code representations in the latent space, 
making the subsequent similarity matching more rational. 
The regularization term can be the KL divergence between the two modes or the KL divergence between 
one mode and the mean of both modes. 
Here, 
we use the KL divergence from query to code, 
expressed as
${KL}\left(q_\theta^{(1)}\left(\vec{z}_q \mid \vec{x}_q\right), q_\theta^{(2)}\left(\vec{z}_{\boldsymbol{c}} \mid \vec{x}_{\boldsymbol{c}}\right)\right)$, 
where $q_\theta^{(1)}\left(\vec{z}_q \mid \vec{x}_q\right)$ and $q_\theta^{(2)}\left(\vec{z}_{\boldsymbol{c}} \mid \vec{x}_{\boldsymbol{c}}\right) $ 
represent the normal distribution consisting of the mean and variance of the query and code. 
The KL divergence from code to query can be obtained by reversing the above formula. 
The average distribution of queries and codes can be obtained by averaging their mean and variance, 
and the mean can also be used as a regularization term in the KL divergence.

\subsubsection{Other Training Methods} 
Besides the pre-training and training techniques introduced above, 
there are also works exploring model training techniques from other perspectives. 
For example, 
meta-learning in \cite{chai2022cross} uses a small verification set to improve model initialization for better performance with small data. 
\cite{wang2022bridging} proposes data enhancement that preserves semantics and uses curriculum learning to control sample weight for better model convergence. 
In zero-shot settings, with no label training process, 
pre-training is crucial for developing a representation ability. 
Hence, 
the input is transformed into the representation space, where the resulting vectors carry semantic information. 
Decisions are made based on the similarity evaluation within this space \cite{guo2022unixcoder}.

\newtcolorbox{RQ3box}{colback=white, colframe = black}
\begin{RQ3box}
  \textbf{\textit{Summary of answers to RQ3: }}
  \begin{itemize}
    \item \textit{Over the past six years, 
the Masked Language Model (MLM) has emerged as the prevailing choice for pre-training tasks in code understanding. 
Despite its apparent simplicity, 
this task has consistently demonstrated remarkable effectiveness and serves as a fundamental cornerstone in achieving comprehensive code comprehension.}
    \item \textit{The training/fine-tuning of code search models predominantly adopt a discriminative model training approach, with pair-wise training being the most prevalent form. }
  \end{itemize}
\end{RQ3box}

\section{Datasets and Evaluation (RQ4)}
Datasets have had a significant impact on the evolution of code search technology. 
They not only serve as a mean to assess the performance of various models, 
but also play a crucial role in addressing real-world challenges faced in code search, 
thus driving the growth of this field. 
In this section, 
we present an overview of 12 code search datasets that have been proposed since 2018 and 8 commonly used metrics for evaluating code search models. 
To provide some guidelines for industrial practitioners, we further discuss how to choose proper code search approaches to fit their needs.

\subsection{Code Search Datasets} 
\begin{figure}[htbp]
	\centering
	\begin{minipage}[c]{0.48\textwidth}
		\centering
		\includegraphics[width=\textwidth]{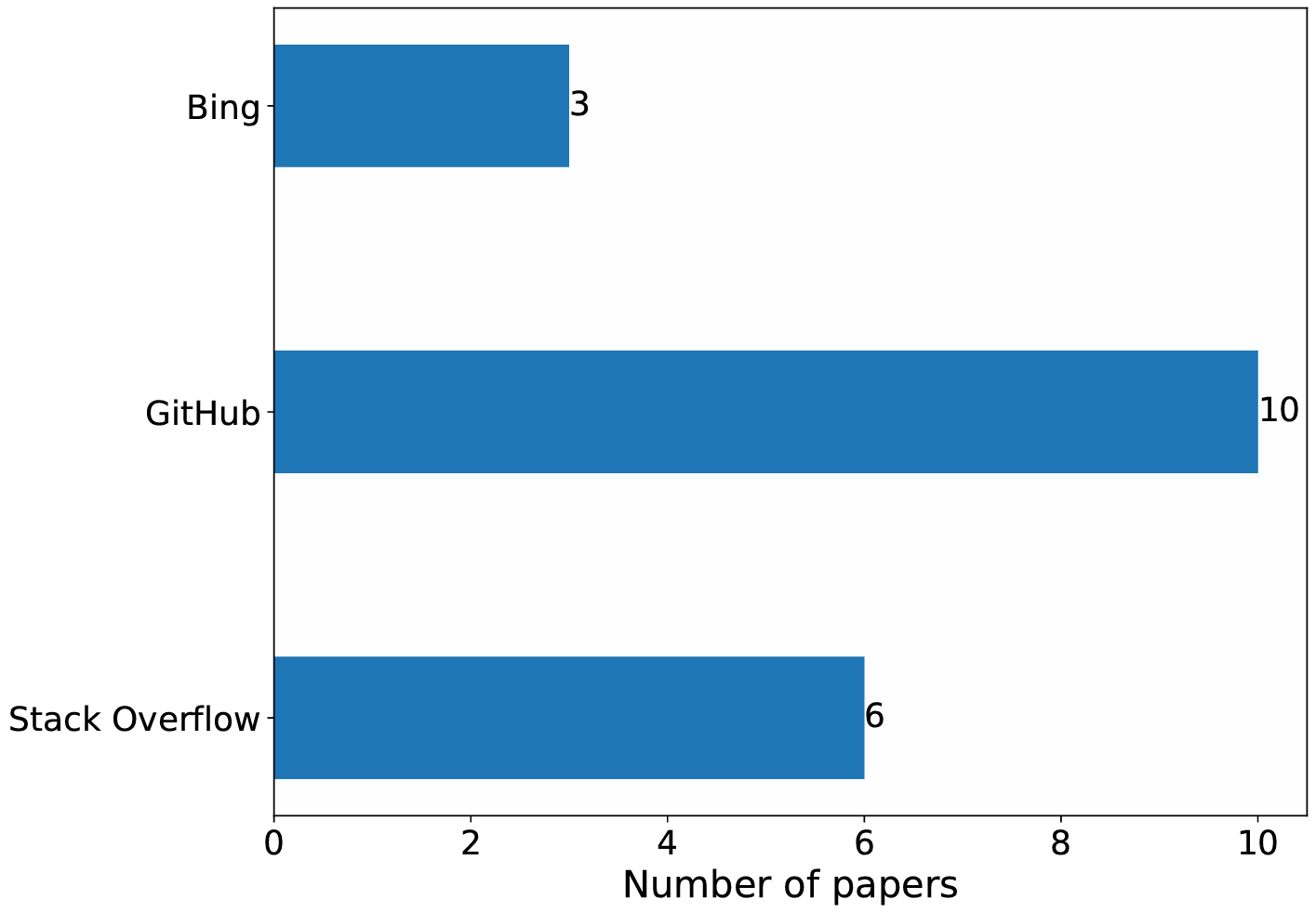}
		\subcaption{Data source}
		\label{fig_E2_1}
	\end{minipage} 
	\begin{minipage}[c]{0.48\textwidth}
		\centering
		\includegraphics[width=\textwidth]{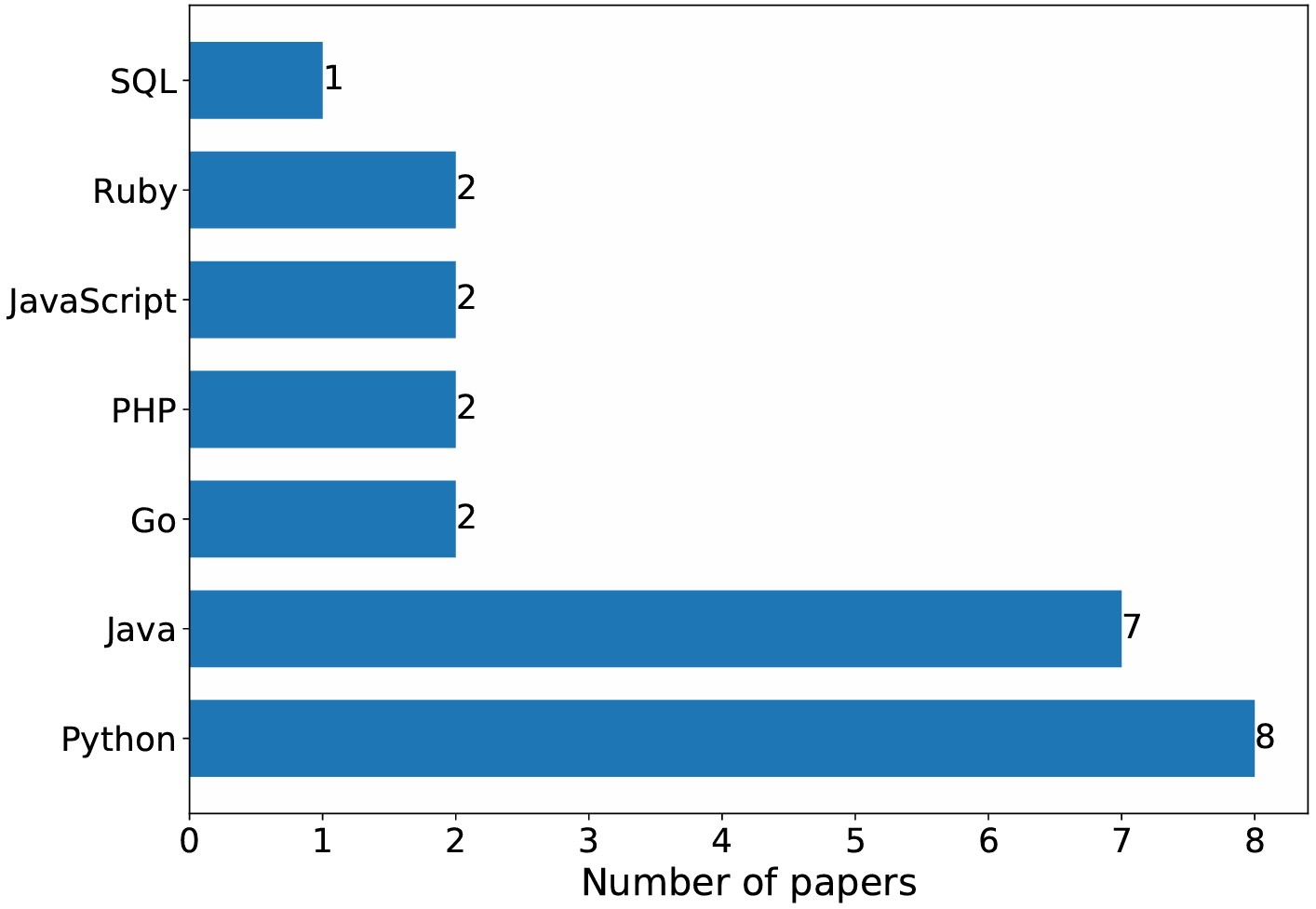}
		\subcaption{Programming language}
		\label{fig_E2_2}
	\end{minipage}
	\caption{Data source and programming language of the datasets.}
	\label{fig_E2}
\end{figure}

% Code search plays a crucial role in software development by providing relevant code snippets in response to natural language queries from the internet or codebase, thereby enhancing developer productivity. 
To drive the advancement of code search, 
various datasets comprising of (text, code) pairs have been introduced. 
This section summarizes the existing datasets on natural language code search research, 
including release year, data sources, programming languages, data statistics, dataset division, data annotation, and acquisition methods, as displayed in Table \ref{table:datasets}. 
In the subsequent section, we give a concise overview of some of the classic corpora outlined in Table \ref{table:datasets}. 
Furthermore,
We conduct meticulous statistical analysis on data sources and programming languages. 
As depicted in Figure \ref{fig_E2}, 
GitHub emerges as the primary source of code search task datasets, 
with Python and Java being the predominant languages of interest for this task.

\begin{table}[H]
  \caption{\textbf{Overview of existing datasets on code search.}}
  \centering
  \resizebox{\textwidth}{!}{
  \begin{tabular}{cccccccc}
  \toprule
  \makecell[c]{Release \\ year}        & Corpus & \makecell[c]{Data \\ sources} & \makecell[c]{Programming \\ languages}                      &      \makecell[c]{Data \\ statistics}                                                          &             \makecell[c]{Data \\ splits}                           &  \makecell[c]{Data \\ annotation}    &     \makecell[c]{Acquisition \\ methods}   \\
  \midrule
  2018  &    StaQC \cite{yao2018staqc}              &              SO               &    \makecell[c]{Python, \\ SQL}                             & \makecell[c]{147,546 Python (question,code) pairs; \\ 119,519 SQL (question,code) pairs}         &  \makecell[c]{training set, \\ dev set, \\ test set}             &  \makecell[c]{manual, \\ automatic}  &  \makecell[c]{https://github.com/\\LittleYUYU/StackOverflow-\\Question-Code-Dataset}          \\
  \midrule
  2018  &    CoNaLa \cite{yin2018learning}              &              SO               &    \makecell[c]{Python, \\ Java}                            & \makecell[c]{42 Python questions, 736 code blocks; \\ 100 Java questions, 434 code blocks}     &  \makecell[c]{training set, \\ dev set, \\ test set}             &  \makecell[c]{manual, \\ automatic}  &   https://conala-corpus.github.io/   \\
  \midrule
  2018  &    Gu et al.  \cite{gu2018deep}            &            GitHub             &                 Java                                        &  18,233,872 Java methods with docstring                                                          &                          training set                                   &             automatic               &  \makecell[c]{https://github.com/\\guxd/deep-code-search}   \\
  \midrule
  2018  &   Java-50  \cite{gu2018deep}              & \makecell[c]{SO, \\ GitHub}   &                 Java                                        &  \makecell[c]{9,950 Java projects;\\ 16,262,602 Java methods;\\ 50 queries}                       &        \makecell[c]{codebase, \\ evaluation set}                    &             manual                   &  \makecell[c]{https://github.com/\\guxd/deep-code-search}   \\
  \midrule
  2019  &  FB-Java  \cite{li2019neural}             & \makecell[c]{SO, \\ GitHub}   &    \makecell[c]{Java}                           &  \makecell[c]{24,549 repositories;\\ 4,716,814 methods;\\ 287 (question,answer) pairs}            &        \makecell[c]{codebase, \\ evaluation set}                    &             manual                   &  \makecell[c]{https://github.com/\\facebookresearch/Neural-Code\\-Search-Evaluation-Dataset}   \\
  \midrule
  2019  &  CSN \cite{husain2019codesearchnet}            &            GitHub             & \makecell[c]{Python,\\ Java, Ruby,\\ Go, PHP,\\ JavaScript} & \makecell[c]{2,326,976 (documentation,function) pairs;\\4,125,470 functions without documentation} &   \makecell[c]{training set,\\ dev set,\\ test set,\\ codebase}  &             automatic               &  \makecell[c]{https://github.com\\/github/CodeSearchNet}  \\
  \midrule
  2019  &  CSN-99 \cite{husain2019codesearchnet}            &    \makecell[c]{Bing, \\ GitHub}    & \makecell[c]{Python,\\ Java, Ruby,\\ Go, PHP,\\ JavaScript} & \makecell[c]{99 queries;\\ 4,026 (query,code) pairs}                                            &                           evaluation set                            &             manual                   &  \makecell[c]{https://github.com\\/github/CodeSearchNet}  \\
  \midrule
  2020  &  SO-DS \cite{heyman2020neural}            & \makecell[c]{SO, \\ GitHub}   &                Python                                       & \makecell[c]{1,113 queries;\\ 12,137 code snippets}                                              &  \makecell[c]{training set, \\ dev set, \\ test set}             &             automatic               &  \makecell[c]{https://github.com/\\nokia/codesearch}  \\
  \midrule
  2020  &  CosBench  \cite{yan2020code}            & \makecell[c]{SO, \\ GitHub}   &                 Java                                        & \makecell[c]{1,000 Java projects;\\ 475,783 Java files;\\ 4,199,769 code snippets;\\ 52 queries}   &        \makecell[c]{codebase, \\ evaluation set}                    &             manual                   &  \makecell[c]{https://github.com/\\BASE-LAB-SJTU/CosBench}  \\
  \midrule
  2020  &  WebQueryTest  \cite{lu2021codexglue}            & \makecell[c]{Bing, \\ GitHub} &                Python                                       & 1,046 (web query, code) pairs                                                                   &                            test set                                     &             manual                   &  \makecell[c]{https://github.com/microsoft/\\CodeXGLUE/tree/main/Text-Code/\\NL-code-search-WebQuery}  \\
  \midrule
  2020  & AdvTest  \cite{lu2021codexglue}             &            GitHub             &                Python                                       & 280,634 (documentation, function) pairs                                                         &  \makecell[c]{training set, \\ dev set, \\ test set}             &             automatic               &  \makecell[c]{https://github.com/microsoft/\\CodeXGLUE/tree/main/Text-Code/\\NL-code-search-Adv}  \\
  \midrule
  2021  & CoSQA  \cite{huang2021cosqa}              & \makecell[c]{Bing, \\ GitHub} &                Python                                       & 20,604 (web query, code) pairs                                                                  &  \makecell[c]{training set, \\ dev set, \\ test set}             &             manual                   &  \makecell[c]{https://github.com/\\Jun-jie-Huang/CoCLR}  \\
  \bottomrule
  \end{tabular}
  }
  \label{table:datasets}
\end{table}

\textbf{StaQC} (\textbf{Sta}ck Overflow \textbf{Q}uestion-\textbf{C}ode pairs) \cite{yao2018staqc}  
is a dataset of (question,code) pairs that has been automatically extracted from Stack Overflow (SO), 
which makes it ideal for predicting whether a code snippet can answer a particular question. 
Stack Overflow is a well-known website where developers can ask programming-related questions, such as ``how to read a file in Python''. 
There are various user-generated solutions available on the site, 
and the ``accepted'' label is used to indicate the quality of these solutions. 
To construct the StaQC dataset, 
Yao et al. \cite{yao2018staqc} filtered posts in the Python and SQL domains on Stack Overflow using tags, 
and used a binary classifier to select posts that had "how-to" type questions. 
They then combined manual labeling and automatic extraction methods to select questions and independent, ``accepted'' solutions from these posts to create (question, code) pairs. 
As a result, they obtained 147,546 Python (question, code) pairs and 119,519 SQL (question, code) pairs.

However, 
collecting questions from Stack Overflow can be a tedious and time-consuming process, 
which limits the quality and quantity of the collected corpus and may pose challenges for systematic comparison of various models. 
With the exponential growth of open-source software projects on GitHub, 
it has become the main source for obtaining code corpus.

\textbf{CSN} (CodeSearchNet) \cite{husain2019codesearchnet} 
is a code search dataset that has been constructed using the open-source projects on GitHub. 
This corpus encompasses 6 different programming languages, including Python, Java, Go, PHP, JavaScript, and Ruby. 
In order to minimize the manual labeling effort, 
Husain et al. \cite{husain2019codesearchnet} replaced natural language queries with documentations, 
forming pairs of (documentation, function) along with code snippets. 
To ensure the corpus's quality, multiple filtering rules were established. 
This involved eliminating (documentation, function) pairs with fewer than 3 documentation tokens, 
functions containing less than 3 lines of code, 
functions with ``test'' in their names, and duplicated functions. 
As a result, 
the CSN corpus comprises a total of 6 million samples, including 2 million (documentation, function) pairs and 4 million functions without paired documents. 
The arrival of the CSN corpus presents exciting opportunities for training large models of code intelligence.

\textbf{AdvTest} \cite{lu2021codexglue} 
is designed for evaluating the generalization capability of code search models. 
It is constructed using the CodeSearchNet's Python corpus. 
To guard against over-fitting, 
Lu et al. \cite{lu2021codexglue} replaced the function and variable names in the test set with special tokens (e.g., ``Func'' in place of a function name). 
This helps prevent the model from over-relying on keyword matching during training and 
decrease the rate of keyword coincidence between query and code at the token level. 

\textbf{CoSQA} \cite{huang2021cosqa} 
is a unique code search dataset that is more representative of real-world code search scenarios compared to other datasets. 
Unlike other datasets that utilize documents, docstrings, or comments as queries, 
CoSQA is based on real user queries collected from Microsoft's Bing search engine logs. 
To construct the CoSQA dataset, 
Huang et al. \cite{huang2021cosqa} filtered queries that did not contain the keyword ``Python'' or showed no code search intent. 
To build (query, code) pairs, 
Huang et al. \cite{huang2021cosqa} utilized CodeBERT to pre-select high-confidence functions from the CodeSearchNet Python corpus for each query. 
Subsequently, the annotators were tasked with determining whether the query was a match for the selected function. 
Finally, they obtained 20,604 (web query, code) pairs that could assist the model in learning the semantic relationship between queries and codes in real-world scenarios.

\begin{table}[H]
  \caption{\textbf{Manually labeled datasets.}}
  \centering
  \resizebox{\textwidth}{!}{
  \begin{tabular}{ccccccc}
  \toprule
  \makecell[c]{Release \\ year}        & Corpus     & number of languages & number of participants   &  type of participant      &     type of query     &   type of codebase     \\
  \midrule
  2018  &    StaQC \cite{yao2018staqc}              &         2           &           4              & undergraduate student     &   question in SO post &   answer in SO post     \\
  \midrule
  2018  &    CoNaLa \cite{yin2018learning}          &         2           &           5              & researcher and programmer &   question in SO post &   answer in SO post     \\
  \midrule
  2018  &   Java-50  \cite{gu2018deep}              &         1           &           2              & programmer                &  question in SO post  &   code in GitHub      \\
  \midrule
  2019  &  FB-Java  \cite{li2019neural}             &         1           &           unknown        & unknown                   &   question in SO post &   answer in SO post     \\
  \midrule
  2019  &  CSN-99 \cite{husain2019codesearchnet}    &         6           &           unknown        & programmer                &   query from Bing     &   code in GitHub  \\
  \midrule
  2020  &  CosBench  \cite{yan2020code}             &         1           &           unknown        & unknown                   &   question in SO post &   answer in SO post and code in GitHub \\
  \midrule
  2020  &  WebQueryTest  \cite{lu2021codexglue}     &         1           &           13             & programmer             &   query from Bing     &  code in GitHub  \\
  \midrule
  2021  & CoSQA  \cite{huang2021cosqa}              &         1           &     more than 100        & programmer                &   query from Bing     &   code in GitHub  \\
  \bottomrule
  \end{tabular}
  }
  \label{table:manually_labeled}
\end{table}

We present a detailed analysis of the manually annotated datasets in Table \ref{table:manually_labeled}. 
Our findings reveal that within these datasets, 
the queries predominantly originate from questions in Stack Overflow (SO) posts, 
closely followed by real queries entered by users in the Bing search engine. 
In comparison to source code comments, 
both of these queries bear a closer resemblance to queries encountered in real search scenarios. 
Notably, the primary source of code in the codebase is the open-source code repository, Github. 
Moreover, code embodies knowledge within the professional realm, 
thereby necessitating the involvement of programmers in the annotation process. 
This factor significantly escalates the costs associated with code annotation.

Based on the analysis above, 
there are three main challenges in the current code search datasets: 
(1) Inconsistency between training data and real user queries. 
The existing datasets primarily consist of (question, code) pairs, (document, code) pairs, and (comment, code) pairs. 
However, there is a discrepancy between the query text input by a user in the search engine and the text found in a question on Stack Overflow or a document/comment in the source code, 
leading to a poor performance of the trained model in real-world scenarios. 
(2) Scarcity of high-quality labeled data. 
Due to the high cost of code labeling, 
the existing datasets are mainly manually labeled in the evaluation set, 
and there is a shortage of a large number of manually labeled training sets, 
restricting the training of supervised learning models. 
(3) Limited data volume. 
The number of training data in existing datasets is limited and currently only reaches a few million, 
which is insufficient for training large-scale code understanding pre-training models. 
Although Markovtsev and Long \cite{markovtsev2018public} selected 182,014 repositories from Github 
to create the first large-scale public code dataset for large-scale programming analysis, 
it is currently not accessible. 
In the future, 
obtaining a large-scale code corpus from Google BigQuery, 
which collects active data on GitHub including complete snapshots of over one million open source repositories and hundreds of millions of code submissions, 
may be explored.

\subsection{Evaluation Metrics}
Code search evaluation datasets typically include carefully selected queries and annotated functions. 
The purpose of evaluating a code search model is to assess its ability to accurately retrieve relevant code snippets from the codebase in response to a given natural language query. 
The evaluation of code search models is largely conducted through automatic methods. 
Common metrics used in recent studies for evaluating code search models include 
Precision, Recall, F1-score, MAP, MRR, Frank, and SuccessRate.

\begin{figure}[htbp]
  \centering
  \includegraphics[width=0.4\linewidth]{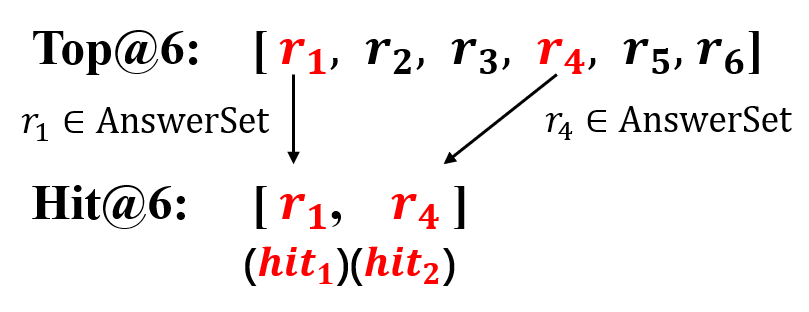}
  \caption{Schematic diagram of $top@k$ and $hit@k$.}
  \label{metric}
\end{figure}

Assume that a given set of queries to be executed, denoted as $Q=\left[q_1, q_2, \ldots, q_n\right]$, 
and each query is marked with its corresponding set of ground-truth answers $G$. 
As depicted in Figure \ref{metric}, 
$top@k$ represents the top-$k$ result sets returned for a particular query, 
and $hit@k$ refers to the answer set among the top-$k$ results that correctly belongs to the ground-truth answer.

\textbf{Precision@k} 
is a metric that shows the relationship between the results of a query and the ground-truth set. 
It measures, on average, how many of the top $k$ results returned by a query belong to the ground-truth set for a query set $Q$. 
The higher the value of $Precision@k$, the stronger the correlation between the returned results and the query. 
It is calculated as follows:
\begin{equation}
  \label{equa26}
  Precision@k =\frac{1}{|Q|} \sum_{i=1}^{|Q|} \frac{\mid  Hit@k \left(q_i\right) \mid}{k}.
\end{equation}

\textbf{Recall@k}  
indicates the average percentage of the ground-truth answer set for each query that is hit. 
The higher the $Recall@k$ value, the more ground-truth answers are retrieved. 
It is calculated as follows:
\begin{equation}
  \label{equa27}
  Recall@k =\frac{1}{|Q|} \sum_{i=1}^{|Q|} \frac{\mid Hit@k \left(q_i\right) \mid}{\left|G_i\right|}.
\end{equation}

\textbf{F1-Score}   
is employed to evaluate the performance of a model when both precision and recall carry equal weight. 
It is calculated as follows:
\begin{equation}
  \label{equa28}
   F1-Score =\frac{2 \cdot  Precision  \cdot  Recall }{ Precision + Recall }.
\end{equation}

\textbf{MAP@k}, 
which stands for Mean Average Precision, reflects the average precision of the rankings produced by all queries. 
It is calculated as follows:
\begin{equation}
  \label{equa29}
   MAP@k =\frac{1}{|Q|} \sum_{i=1}^{|Q|} \frac{1}{m} \sum_{j=1}^m \frac{j}{{rank}\left( { hit }_j, { Top@k }\left(q_i\right)\right)},
\end{equation} 
where $m$ is $|Hit@k|$, $hit_j$ is an element in $Hit@k$, 
and ${rank}(e, l)$ is the rank (i.e., index) of element $e$ in list $l$. 
When $\left|Hit @ k\left(q_{i}\right)\right|=0$, 
$hit_j$ does not exist, and the average precision of query $q$ is 0. 
It is evident that a larger $MAP@k$ value signifies that a greater number of answers that hit the ground-truth are present in the top-$k$ results returned.

\textbf{MRR@k}, 
which stands for Mean Reciprocal Rank, indicates the average of the reciprocals of the rankings of all search results. 
It is calculated as follows:
\begin{equation}
  \label{equa30}
  MRR@k=\frac{1}{|Q|} \sum_{i=1}^{|Q|} \frac{1}{{rank}\left( { hit }_1, {Top@k}\left(q_i\right)\right)}.
\end{equation}
When $\left|Hit @ k\left(q_{i}\right)\right|=0$, 
the reciprocal of the ranking is 0. 
Typically, 
only the first answer that hits the ground-truth is taken into account when calculating $MRR@k$. 
The greater the $MRR@k$ value, the higher the ranking of the answer that hits the ground-truth in $Top@k$.

\textbf{Frank@k}  
represents the average rank of the first hit answer across all queries. 
It is calculated as follows:
\begin{equation}
  \label{equa31}
  Frank@k=\frac{1}{|Q|} \sum_{i=1}^{|Q|} {rank}\left({hit}_1\left(q_i\right), {Top@ k}\left(q_i\right)\right).
\end{equation}
It is evident that a smaller $Frank@k$ value corresponds to a higher ranking of the ground-truth answer in the search results.

\textbf{SuccessRate@k}  
indicates the proportion of queries for which there are more than one ground-truth answers among the top-$k$ results returned.  
It is calculated as follows:
\begin{equation}
  \label{equa32}
   SuccessRate@k =\frac{1}{|Q|} \sum_{i=1}^{|Q|} \tau\left({Frank}_{q_i} \leq k\right),
\end{equation}
where $\tau$ is an indicator function 
that returns 1 when the ranking of the hit answer for query $q_i$ is less than $k$ and 0 otherwise. 
$SuccessRate@k$ is a crucial evaluation metric because an effective code search engine should enable software developers to find the desired code snippets by examining fewer results.

\textbf{NDCG}, which stands for Normalized Discounted Cumulative Gain, 
serves as a crucial metric to quantify the similarity between the ranking of candidate code fragments and the ideal ranking, 
placing significant emphasis on the overall ranking order among all candidate results. 
It is calculated as follows:
\begin{equation}
  \label{equa33}
  NDCG=\frac{1}{|Q|} \sum_{i=1}^k \frac{2^{r_i}-1}{\log _2(i+1)},
\end{equation}
where $r_i$ is the relevance score of the top-$k$ results at position $i$.

\begin{figure}[htbp]
	\centering
	\begin{minipage}[c]{0.48\textwidth}
		\centering
		\includegraphics[width=\textwidth]{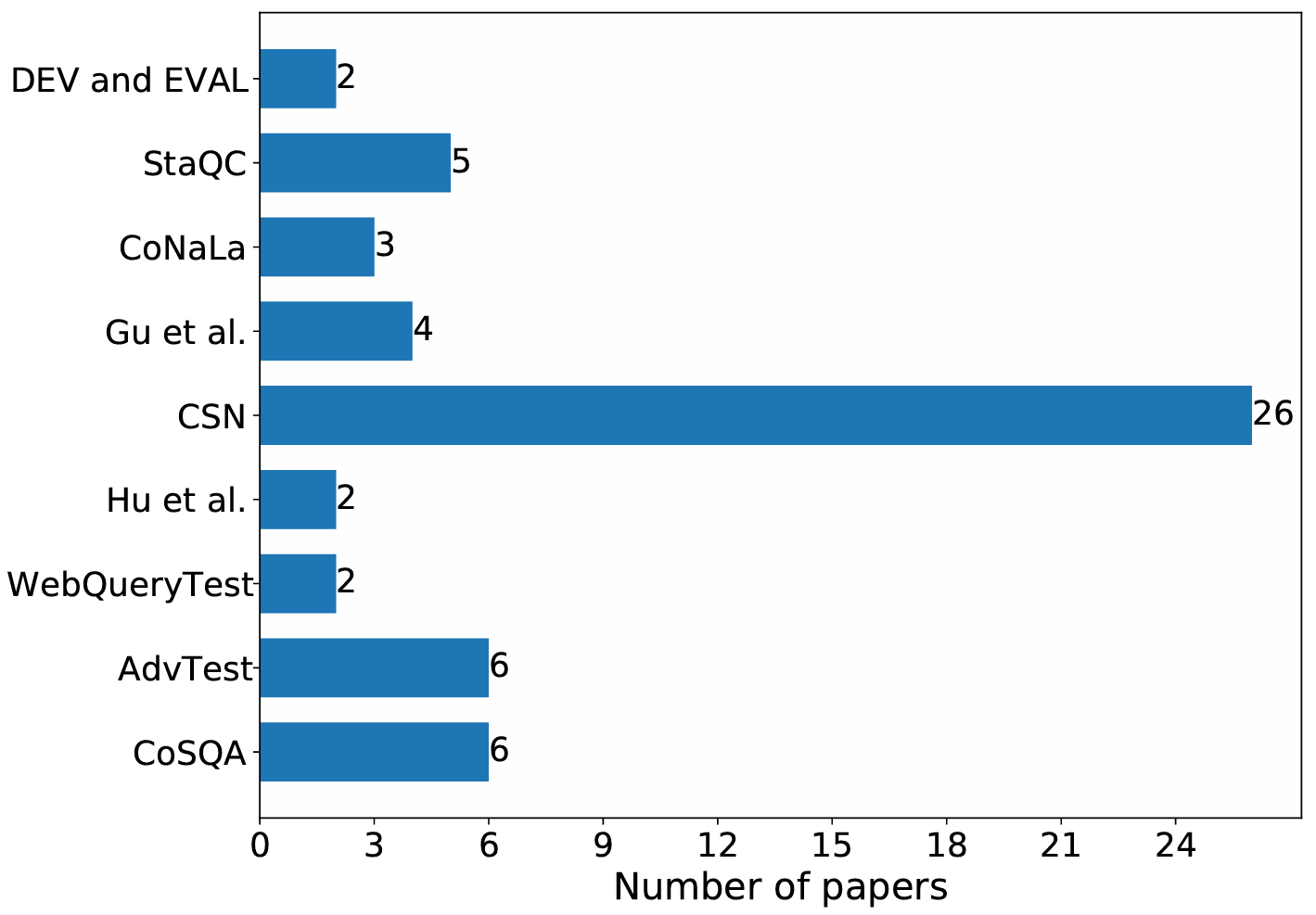}
		\subcaption{Datasets adopted by at least two papers.}
		\label{fig_E3_1}
	\end{minipage} 
	\begin{minipage}[c]{0.48\textwidth}
		\centering
		\includegraphics[width=\textwidth]{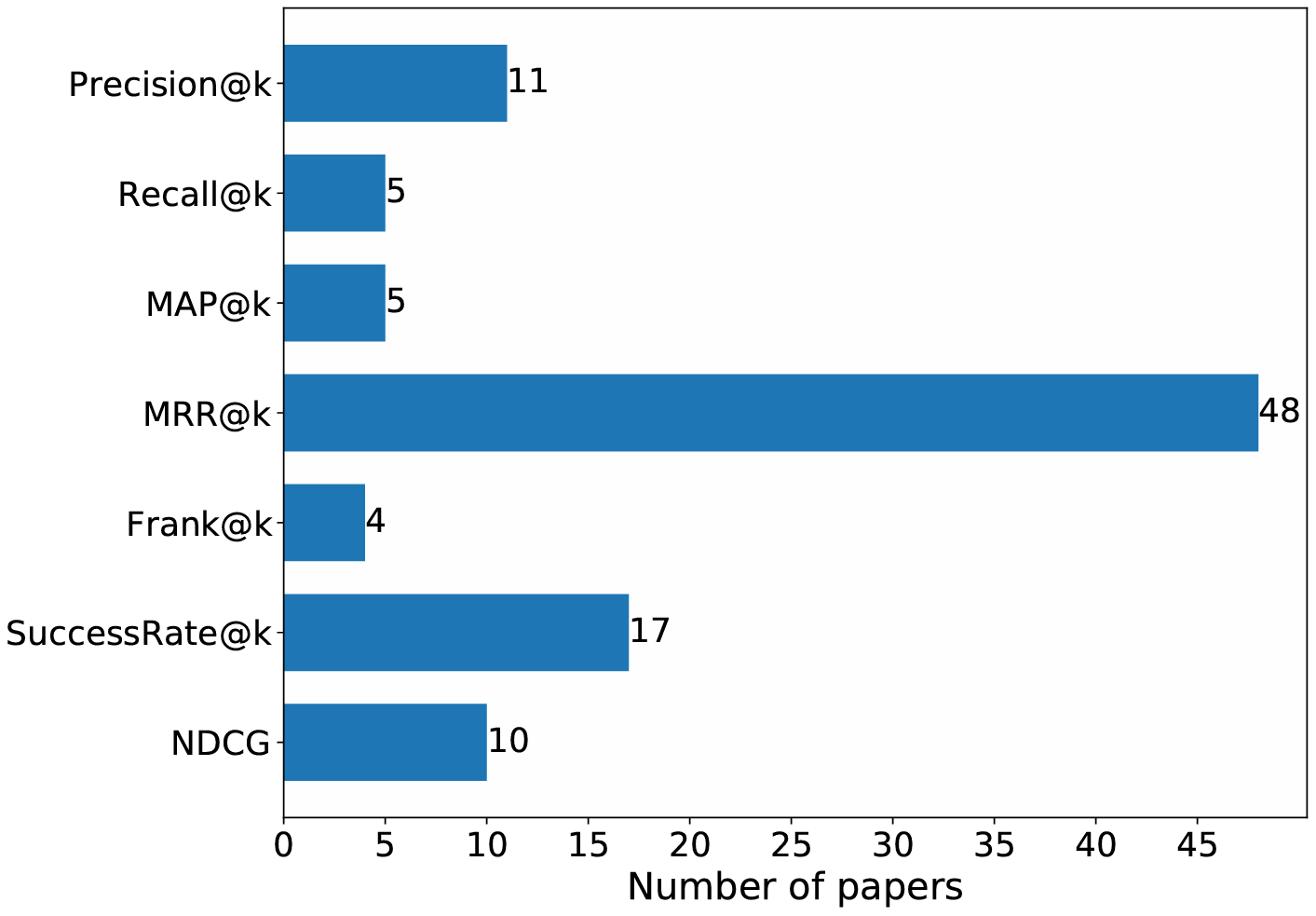}
		\subcaption{Metrics adopted by at least two papers.}
		\label{fig_E3_2}
	\end{minipage}
	\caption{Datasets and metrics adopted by at least two papers.}
	\label{fig_E3}
\end{figure}

Table \ref{table:models_and_metrics} provides an in-depth analysis of 53 deep learning-based code search models, 
encompassing the evaluated datasets, the baseline model utilized for comparison, and the selected evaluation metrics. 
Figure \ref{fig_E3} highlights that the CSN dataset holds prominence in the code search task, 
while the most widely adopted evaluation metric is MRR@k, with a utilization rate of 90.6\%.

\begin{table}[htbp]
  \caption{\textbf{Deep code search models and their performance metrics.}}
  \centering
  \scalebox{0.5}{
  \begin{tabular}{ccccc}
  \toprule
  Year & Models                                      & Dataset                                    &              Baselines               & Metric             \\
  \midrule
  2018 & CODEnn \cite{gu2018deep}                    & Gu et al. \cite{gu2018deep}                &   CodeHow \cite{lv2015codehow}       &  Frank@k, Precision@k, MRR@k, SuccessRate@k                  \\
  \midrule
  2018 & NCS \cite{sachdev2018retrieval}             & Sachdev et al. \cite{sachdev2018retrieval} &   TF-IDF, BM25                       &  Precision@k     \\
  \midrule
  2018 & NLP2API \cite{rahman2018effective}          &  Rahman and Roy \cite{rahman2018effective} &   Rahman and Roy\cite{rahman2018effective}        &  \makecell[c]{Precision@k, MRR@K,\\ MAP@K, Recall@K, Frank@k}           \\
  \midrule
  2018 & COCABU \cite{sirres2018augmenting}          &  Sirres et al. \cite{sirres2018augmenting} &  Codota, OpenHub                     & Precision@k, MRR@k          \\
  \midrule
  2018 & Zhang et al. \cite{zhang2017expanding}      & Zhang et al. \cite{zhang2017expanding}     &  Dice, Rocchio, RSV                  & MRR@k, SuccessRate@k          \\
  \midrule
  2019 & UNIF \cite{cambronero2019deep}              & Gu et al. \cite{gu2018deep}                & CODEnn \cite{gu2018deep}, NCS \cite{sachdev2018retrieval}  & SuccessRate@k     \\
  \midrule
  2019 & MMAN \cite{wan2019multi}                    &  Wan et al. \cite{wan2019multi}                   & CodeHow \cite{lv2015codehow}, CODEnn \cite{gu2018deep}     & MRR@k, SuccessRate@k  \\
  \midrule
  2019 & RACK \cite{rahman2019automatic}             &  Rahman \cite{rahman2019automatic}         & NL Keywords \cite{rahman2019automatic}                     & \makecell[c]{Precision@k, MRR@K, MAP@K,\\ Recall@K, NDCG, Frank@k}           \\
  \midrule
  2019 & Rahman \cite{rahman2019supporting}          &  Rahman \cite{rahman2019supporting}        &  -                                   & Hit@K, MAP@k, MRR@k, Recall@k, Frank@k  \\
  \midrule
  2019 & NQE \cite{liu2019neural}                    &  Liu et al. \cite{liu2019neural}                  & BM25, NCS \cite{sachdev2018retrieval} &  Precision@k, MRR@k    \\
  \midrule
  2019 & QESC \cite{huang2019deep}                   &  Huang et al. \cite{huang2019deep}                &  CodeHow \cite{lv2015codehow}, QECK \cite{nie2016query}    & Precision@k, NDCG           \\
  \midrule
  2019 & CoaCor \cite{yao2019coacor}                 & StaQC \cite{yao2018staqc}, DEV and EVAL \cite{iyer2016summarizing}     &    CODEnn \cite{gu2018deep}, CODE-NN \cite{iyer2016summarizing}      &     MRR@k               \\
  \midrule
  2019 & Wu and Yang \cite{wu2019code}               &  Wu and Yang \cite{wu2019code}             &    CodeHow \cite{lv2015codehow}          & Precision@k, NDCG           \\
  \midrule
  2020 & OCoR \cite{zhu2020ocor}                     &  StaQC \cite{yao2018staqc}, DEV and EVAL \cite{iyer2016summarizing}      &  CODEnn \cite{gu2018deep}, CODE-NN \cite{iyer2016summarizing}, CoaCor \cite{yao2019coacor}                &    MRR@k                \\
  \midrule
  2020 & CodeBERT \cite{feng2020codebert}            & CSN \cite{husain2019codesearchnet}         &  NBoW, 1D-CNN, biRNN, SelfAtt, RoBERTa(code)                                &   MRR@k                 \\
  \midrule
  2020 & AdaCS \cite{ling2020adaptive}               &  Ling et al. \cite{ling2020adaptive}    & CodeHow \cite{lv2015codehow}, CODEnn \cite{gu2018deep}, BVAE \cite{chen2018neural}   &   Hit@K, MRR@k             \\
  \midrule
  2020 & MP-CAT \cite{haldar2020multi}               & CoNaLa \cite{yin2018learning}              &   CT              &   Recall@k, MRR@k                 \\
  \midrule
  2020 & ${TranS}^3$ \cite{wang2020trans}            & Barone and Sennrich \cite{barone2017parallel}     &  CODEnn \cite{gu2018deep}, CoaCor \cite{yao2019coacor}, Hybrid-DeepCom \cite{hu2020deep}, AutoSum \cite{wan2018improving}   &  MRR@k, NDCG, SuccessRate@k     \\
  \midrule
  2020 & Zhao and Sun \cite{zhao2020adversarial}     & StaQC \cite{yao2018staqc} & CODEnn \cite{gu2018deep}, CoaCor \cite{yao2019coacor}     &  MAP@k, NDCG                  \\
  \midrule
  2020 & CO3 \cite{ye2020leveraging}                 & StaQC \cite{yao2018staqc}  &  CODEnn \cite{gu2018deep}, CoaCor \cite{yao2019coacor}   &  MRR@k, NDCG         \\
  \midrule
  2021 & GraphCodeBERT \cite{guo2020graphcodebert}   & CSN \cite{husain2019codesearchnet}    &  NBoW, 1D-CNN, biRNN, SelfAtt, RoBERTa(code), CodeBERT \cite{feng2020codebert}  &   MRR@k                 \\
  \midrule
  2021 & DGMS \cite{ling2021deep}                    &  FB-Java \cite{li2019neural},  CSN-Python \cite{husain2019codesearchnet}      & NBoW, 1D-CNN, biRNN, SelfAtt, CODEnn \cite{gu2018deep}, UNIF \cite{cambronero2019deep}, CAT \cite{haldar2020multi}    &  MRR@k, SuccessRate@k    \\
  \midrule
  2021 & CoCLR \cite{huang2021cosqa}                 &  CoSQA  \cite{huang2021cosqa}, WebQueryTest \cite{lu2021codexglue}       &   RoBERTa \cite{liu2019roberta}, CodeBERT \cite{feng2020codebert}        &   MRR@k                 \\
  \midrule
  2021 & SEQUER \cite{cao2021automated}              & Cao et al. \cite{cao2021automated}        &   seq2seq \cite{sutskever2014sequence}                                   &   MRR@k                 \\
  \midrule
  2021 & DOBF \cite{lachaux2021dobf}                 & CSN-Python \cite{husain2019codesearchnet}      &  CodeBERT \cite{feng2020codebert}, GraphCodeBERT \cite{guo2020graphcodebert}               &  MRR@k             \\
  \midrule
  2021 & CRaDLe \cite{gu2021cradle}                  & CSN \cite{husain2019codesearchnet}  &  \makecell[c]{CODEnn \cite{gu2018deep}, UNIF \cite{cambronero2019deep},\\ NBoW, 1D-CNN, biRNN, SelfAtt, ConvSelfAtt }  &  MRR@k, SuccessRate@k                  \\
  \midrule
  2021 & Corder \cite{bui2021self}                   & Gu et al. \cite{gu2018deep} &   NBoW, biRNN, SelfAtt, TBCNN \cite{mou2016convolutional}, Code2vec \cite{alon2019code2vec}   &  Precision@k, MRR@k                  \\
  \midrule
  2021 & Gu et al. \cite{gu2021multimodal}           & CSN \cite{husain2019codesearchnet} &  NBoW, 1D-CNN, biRNN, SelfAtt                        &  MRR@k, NDCG             \\
  \midrule
  2021 & TabCS \cite{xu2021two}                      &  Hu et al. \cite{hu2020deep}, CSN \cite{husain2019codesearchnet}            &    CODEnn \cite{gu2018deep}, CARLCS-CNN \cite{shuai2020improving}, CARLCS-TS \cite{shuai2020improving}, UNIF \cite{cambronero2019deep}  &  MRR@k, SuccessRate@k         \\
  \midrule
  2021 & MuCoS \cite{du2021single}                   &  CSN \cite{husain2019codesearchnet}  & \makecell[c]{NBoW, 1D-CNN, biRNN, SelfAtt, ConvSelfAtt,\\ CODEnn \cite{gu2018deep}, CodeBERT \cite{feng2020codebert}   }        &  MRR@k, SuccessRate@k    \\
  \midrule
  2021 & SynCoBERT \cite{wang2021syncobert}          &  CSN \cite{husain2019codesearchnet}, AdvTest \cite{lu2021codexglue}        & \makecell[c]{ NBow, CNN, BiRNN, SelfAttn, RoBERTa \cite{liu2019roberta},\\ RoBERTa(code), CodeBERT \cite{feng2020codebert}, GraphCodeBERT \cite{guo2020graphcodebert} }    &    MRR@k                \\
  \midrule
  2022 & TranCS \cite{sun2022code}                   & CSN-Java \cite{husain2019codesearchnet}   &  CODEnn \cite{gu2018deep}, MMAN \cite{wan2019multi}    &  MRR@k, SuccessRate@k  \\
  \midrule
  2022 & Wang et al. \cite{wang2022bridging}         & CSN \cite{husain2019codesearchnet}        &  \makecell[c]{NBoW, 1D-CNN, biRNN, SelfAtt, RoBERTa \cite{liu2019roberta},\\ RoBERTa(code), CodeBERT \cite{feng2020codebert}, GraphCodeBERT \cite{guo2020graphcodebert} }  &    MRR@k                \\
  \midrule
  2022 & CodeRetriever \cite{li2022coderetriever}    & \makecell[c]{CSN \cite{husain2019codesearchnet}, AdvTest \cite{lu2021codexglue}, CoSQA \cite{huang2021cosqa},\\ CoNaLa \cite{yin2018learning}, SO-DS \cite{heyman2020neural}, StaQC \cite{yao2018staqc} }        &   CodeBERT \cite{feng2020codebert}, GraphCodeBERT \cite{guo2020graphcodebert}, SynCoBERT \cite{wang2021syncobert}            &    MRR@k                \\
  \midrule
  2022 & CDCS \cite{chai2022cross}                   &  \makecell[c]{ CSN-Python \cite{husain2019codesearchnet}, CSN-Java \cite{husain2019codesearchnet},\\ Solidity and SQL \cite{yang2021multi}  }   &  Roberta \cite{liu2019roberta}, CodeBERT \cite{feng2020codebert}     &  MRR@k, SuccessRate@k      \\
  \midrule
  2022 & CSRS \cite{cheng2022csrs}                   &  Gu et al. \cite{gu2018deep} &  CODEnn \cite{gu2018deep}, CARLCS-CNN \cite{shuai2020improving}  & Recall@k, MRR@k, NDCG    \\
  % \midrule
  % 2022 & cpt-code \cite{neelakantan2022text}         & CSN \cite{husain2019codesearchnet} &   CodeBERT \cite{feng2020codebert}, GraphCodeBERT \cite{guo2020graphcodebert}     &     MRR               \\
  \midrule
  2022 & ${NS}^3$ \cite{arakelyan2022ns3}            & CSN \cite{husain2019codesearchnet}, CoSQA \cite{huang2021cosqa} & \makecell[c]{BM25, RoBERTa(code), CuBERT \cite{kanade2020learning},\\ CodeBERT \cite{feng2020codebert}, GraphCodeBERT \cite{guo2020graphcodebert} }   &   MRR@k, Precision@k                 \\
  \midrule
  2022 & CSSAM \cite{hu2022cssam}                    &  Hu et al. \cite{hu2020deep}, CSN \cite{husain2019codesearchnet}  & CodeHow \cite{lv2015codehow}, CODEnn \cite{gu2018deep}, MP-CAT \cite{haldar2020multi}, TabCS \cite{xu2021two}    &  MRR@k, SuccessRate@k, NDCG     \\
  \midrule
  2022 & CTBERT \cite{han2022towards}                & CSN \cite{husain2019codesearchnet}, AdvTest \cite{lu2021codexglue}   & CodeBERT \cite{feng2020codebert}, GraphCodeBERT \cite{guo2020graphcodebert}, SynCoBERT \cite{wang2021syncobert}   &   MRR@k                 \\
  \midrule
  2022 & QueCos \cite{wang2022enriching}             & \makecell[c]{CSN-Python \cite{husain2019codesearchnet}, CSN-Java \cite{husain2019codesearchnet},\\ Wang et al. \cite{wang2022enriching} }    &  CODEnn \cite{gu2018deep}, UNIF \cite{cambronero2019deep}, OCoR \cite{zhu2020ocor}                       & MRR@k, SuccessRate@k       \\
  \midrule
  2022 & ZaCQ \cite{eberhart2022generating}          & CSN \cite{husain2019codesearchnet}  &  V-DO, KW       & MRR@k, MAP@k, NDCG                   \\
  \midrule
  2022 & G2SC \cite{shi2022better}                   & CSN \cite{husain2019codesearchnet}    &  CODEnn \cite{gu2018deep}, MMAN \cite{wan2019multi}, CodeBERT \cite{feng2020codebert}, GraphCodeBERT \cite{guo2020graphcodebert}      &   MRR@k                 \\
  \midrule
  2022 & SPT-Code \cite{niu2022spt}                  & CSN \cite{husain2019codesearchnet}    &  CNN, Bi-GRU, SelfAtt, CodeBERT \cite{feng2020codebert}, GraphCodeBERT \cite{guo2020graphcodebert}      &   MRR@k                 \\
  \midrule
  2022 & CODE-MVP \cite{wang2022code}                & AdvTest \cite{lu2021codexglue}, CoNaLa \cite{yin2018learning}, CoSQA \cite{huang2021cosqa}    &  \makecell[c]{RoBERTa \cite{liu2019roberta}, CodeBERT \cite{feng2020codebert},\\ GraphCodeBERT \cite{guo2020graphcodebert}, SynCoBERT \cite{wang2021syncobert}  }   &   MRR@k                 \\
  \midrule
  2022 & UniXcoder \cite{guo2022unixcoder}           & CSN \cite{husain2019codesearchnet}, AdvTest \cite{lu2021codexglue}, CoSQA \cite{huang2021cosqa}   &  \makecell[c]{RoBERTa \cite{liu2019roberta}, CodeBERT \cite{feng2020codebert},\\ GraphCodeBERT \cite{guo2020graphcodebert}, SynCoBERT \cite{wang2021syncobert}  } &    MRR@k                \\
  \midrule
  2022 & Li et al. \cite{li2022exploring}            & CSN \cite{husain2019codesearchnet}     & \makecell[c]{RoBERTa(code), CodeBERT \cite{feng2020codebert},\\ GraphCodeBERT \cite{guo2020graphcodebert}, UniXCoder \cite{guo2022unixcoder}  }  &  MRR@k                  \\
  \midrule
  2022 & SCodeR \cite{li2022soft}                    & CSN \cite{husain2019codesearchnet}, AdvTest \cite{lu2021codexglue}, CoSQA \cite{huang2021cosqa}  & \makecell[c]{ CodeBERT \cite{feng2020codebert}, GraphCodeBERT \cite{guo2020graphcodebert}, SyncoBERT \cite{wang2021syncobert},\\ CodeRetriever \cite{li2022coderetriever}, Code-MVP \cite{wang2022code}, UniXcoder \cite{guo2022unixcoder}  }  &     MRR@k               \\
  \midrule
  2023 & Salza et al. \cite{salza2022effectiveness}  & Salza et al. \cite{salza2022effectiveness}    &  LUCENE, CODEnn \cite{gu2018deep}            &  MRR@k, SuccessRate@k   \\
  \midrule
  2023 & deGraphCS \cite{zeng2023degraphcs}          & Zeng et al. \cite{zeng2023degraphcs}      &  CODEnn \cite{gu2018deep}, UNIF \cite{cambronero2019deep}, MMAN \cite{wan2019multi}    &  MRR@k, SuccessRate@k    \\
  \midrule
  2023 & GraphSearchNet \cite{liu2023graphsearchnet} & CSN-Python \cite{husain2019codesearchnet}, CSN-Java \cite{husain2019codesearchnet}  &  \makecell[c]{NBoW, 1D-CNN, biRNN, SelfAtt, UNIF \cite{cambronero2019deep},\\ CODEnn \cite{gu2018deep}, CARLCS-CNN \cite{shuai2020improving}, TabCS \cite{xu2021two}, Coacor \cite{yao2019coacor}   }   &  MRR@k, NDCG, SuccessRate@k  \\
  \midrule
  2023 & MulCS \cite{ma2023mulcs}                    & CSN \cite{husain2019codesearchnet}, C dataset \cite{ma2023mulcs}   & CODEnn \cite{gu2018deep}, TabCS \cite{xu2021two}, NBoW, 1D-CNN, biRNN, SelfAtt   &  MRR@k, SuccessRate@k    \\
  \midrule
  2023 & KeyDAC \cite{park2023contrastive}           & CoSQA \cite{huang2021cosqa}, WebQueryTest \cite{lu2021codexglue}    &  CoCLR \cite{huang2021cosqa}     &   MRR@k                 \\
  \midrule
  2023 & TOSS \cite{hu2023revisiting}                &  CSN \cite{husain2019codesearchnet}  & \makecell[c]{Jaccard, BOW, TFIDF, BM25, CODEnn \cite{gu2018deep}, \\ CodeBERT \cite{feng2020codebert}, GraphCodeBERT \cite{guo2020graphcodebert}, CoCLR \cite{huang2021cosqa}}   &   MRR@k, Precision@k                 \\
  \bottomrule
  \end{tabular}
  }
\label{table:models_and_metrics}
\end{table}

\subsection{Usage Scenarios}
The rapid evolution of code search technology has presented the industry with a multitude of fresh opportunities.
An interesting and important question for industrial practitioners is how to choose proper code search approaches to fit their needs. 
This section discusses this question from three aspects: effectiveness, efficiency, and multi-language.

\textbf{Effectiveness.} 
By placing a heightened emphasis on the accuracy of code search, 
industrial practitioners can leverage a suite of powerful techniques. 
Methods such as RACK \cite{rahman2019automatic}, NQE \cite{liu2019neural}, SEQUER \cite{cao2021automated}, and ZaCQ \cite{eberhart2022generating} offer effective means to reconstruct queries, 
aiding in the clarification of user search intent. 
To deepen the understanding of code semantics, 
practitioners can employ methods such as MP-CAT \cite{haldar2020multi}, CRaDLe \cite{gu2021cradle}, GraphCodeBERT \cite{guo2020graphcodebert}, TabCS \cite{xu2021two}, SynCoBERT \cite{wang2021syncobert}, G2SC \cite{shi2022better}, SPT-Code \cite{niu2022spt}, and UniXcoder \cite{guo2022unixcoder}. 
Furthermore, methods such as MP-CAT \cite{haldar2020multi}, TabCS \cite{xu2021two}, CSRS \cite{cheng2022csrs}, ${NS}^3$ \cite{arakelyan2022ns3}, and CSSAM \cite{hu2022cssam} enable fine-grained interaction between query and code, 
effectively bridging the semantic gap that exists between them. 
Considering the pivotal role of the training corpus in determining the performance of neural code search \cite{sun2022importance}, 
it is imperative for practitioners to prioritize the quality of their data. 
By meticulously cleaning the training corpus, 
practitioners can attain a high-quality dataset that fosters the establishment of a precise mapping from natural language to programming language. 
Furthermore, 
practitioners can leverage CodeRetriever \cite{li2022coderetriever} to specifically concentrate on examining the semantic distinctions between query and code via Contrastive Learning. 
This approach facilitates a detailed analysis of the nuanced differences, 
thereby enhancing the overall search accuracy. 
Additionally, 
the recall and rerank framework offered by TOSS \cite{hu2023revisiting} presents an invaluable opportunity to augment the accuracy of code search. 
By integrating this framework, practitioners can achieve further improvements in the accuracy of their products.

\textbf{Efficiency.} 
By prioritizing search efficiency, 
industrial practitioners can leverage the offline calculation of code fragment embeddings within their codebases. 
This approach offers a time-saving alternative to the online calculation of vector similarity. 
Moreover, 
the utilization of CoSHC \cite{gu2022accelerating}, a code search acceleration method based on deep hashing and code classification, 
enables efficient code search while accepting a minor trade-off in precision.

\textbf{Multi-language.} 
By emphasizing the generalization of code search tools across multiple programming languages, 
industrial practitioners can use MulCS \cite{ma2023mulcs} to break down the semantic barriers between different programming languages,
which uses a unified data structure to represent multiple programming languages.

\newtcolorbox{RQ4box}{colback=white, colframe = black}
\begin{RQ4box}
  \textbf{\textit{Summary of answers to RQ4: }}
  \begin{itemize}
    \item \textit{The primary source of the code in the code search datasets originates from the open-source code repository, Github. 
On the other hand, Stack Overflow (SO) serves as the main source of queries these the datasets.}
    \item \textit{Python and Java are the two most concerned programming languages for code search tasks.} 
    \item \textit{For nearly 6 years, the CSN dataset has dominated the evaluation of code search tasks.}
    \item \textit{The prevalent evaluation metrics employed in code search encompass Precision@k, Recall@k, F1-score, MAP@k, MRR@k, Frank@k, and SuccessRate@k, 
with MRR@k widely utilized indicator among them.}
  \end{itemize}
\end{RQ4box}

\section{Open Challenges and Future Directions (RQ5)}
This section addresses the outstanding challenges in the field of code search and identifies potential avenues for future advancements.

\begin{enumerate}[(1)]
  \item \textbf{Understanding code structure}. 
The information contained in code can be divided into two parts: 
structured information determined by keywords that define the code's function, 
and auxiliary information reflected by identifiers such as function and variable names, 
which indicate a software developer's generalization of code functions. 
However, 
studies have shown that anonymizing function and variable names can lead to a significant drop in model performance \cite{guo2022unixcoder}, 
indicating that current methods rely too heavily on identifier information for inference. 
Identifier content serves as a convenient shortcut for code intelligence tasks, 
but an overreliance on it may hinder the model's ability to learn the code functions defined by structured information, 
thereby hindering its overall understanding of code structure. 
The challenge is to design a strategy that allows the model to better focus on and understand the code's structural information, 
and to strike a balance between the structural information defined by keywords and the auxiliary information provided by identifiers during both training and inference. 
This is a direction worthy of further research. 
First steps toward that goal have been taken, e.g., by Li et al. \cite{li2022soft} and Wang et al. \cite{wang2022code}.
  \item \textbf{Graph pre-training technology}. 
Code can be analyzed and represented as a graph data structure (e.g., Data Flow Graph \cite{guo2020graphcodebert} and Variable-based Flow Graph \cite{zeng2023degraphcs}). 
Compared to sequential structures, using a graph structure to represent code provides a larger amount of information and higher accuracy. 
We believe that this graph structure encompasses valuable insights that can lead to breakthroughs in understanding the meaning of code.
However, 
based on the results of existing research, methods utilizing Transformer-encoded sequences \cite{guo2022unixcoder, wang2021syncobert, li2022coderetriever, wang2022code} 
continue to maintain their lead over those based on graph neural network approaches for graph encoding \cite{zeng2023degraphcs, ling2021deep, ma2023mulcs, liu2023graphsearchnet}. 
The remarkable success of the Transformer-based methods can be largely attributed to the substantial improvement in model capacity achieved through pre-training techniques. 
We believe that the potential of graph-based methods in code intelligence has not been fully explored yet. 
A valuable research direction lies in exploring how to incorporate graph pre-training into code intelligence, 
aiming to fully unleash the performance of graph-based models.
  \item \textbf{Robust optimization}. 
Research has indicated that current code intelligence methods are susceptible to adversarial attacks \cite{li2022semantic}. 
Perturbing identifiers significantly decreases the model's performance, highlighting its lack of robustness. 
Efforts have been made to enhance robustness through adversarial training \cite{wan2022you}, but further studies are required.  
At the same time, it is crucial to establish a standard evaluation method or dataset to evaluate the robustness of these models.
  \item \textbf{Interpretability research}. 
Like other deep learning models, 
code search models are trained and operate in a relatively black-box manner. 
Interpretability research aims to understand the reasons behind a model's decisions, 
improve our understanding of the model's inference process, 
and identify the information that the model considers important.  
This research is crucial as it can boost user confidence in the model and, 
in case of errors, enable the identification of the cause and development of a solution in a timely manner. 
Karmakar and Robbes \cite{karmakar2021pre} and Wan et al. \cite{wan2022they} propose promising first steps into this direction.
  \item \textbf{Comprehensive evaluation}. 
Presently, 
the most efficient approach in the field of code search is the pre-training \& fine-tuning paradigm 
(e.g., CodeBERT \cite{feng2020codebert}, GraphCodeBERT \cite{guo2020graphcodebert}, Syncobert \cite{wang2021syncobert}, and UniXcoder \cite{guo2022unixcoder}). 
This paradigm encompasses several crucial design elements, 
including data preprocessing strategy, tokenizer, pre-training task, pre-training model \& training parameters, fine-tuning model \& training parameters, and negative sample construction strategy. 
However, there is a shortage of analytical experiments on these key elements that can guide practitioners in choosing the most effective training strategies for code intelligence. 
For example, 
Guo et al. \cite{guo2020graphcodebert} adopt the parameters of CodeBERT \cite{feng2020codebert} to initialize the GraphCodeBERT model and proceed with continue pre-training on the CSN dataset using tasks like MLM. 
However, the performance improvement has not been validated to determine whether it is a result of additional iterations on the MLM task.
Investigating the significance of these design elements will offer more direction to practitioners 
and also provide new insights for future research by uncovering the unique properties of code intelligence.
  \item \textbf{High-quality datasets}. 
At present, 
the most widely used datasets in code search come from three sources: 
crawling from Github (e.g., CSN \cite{husain2019codesearchnet}), which consists of code comments as queries and the rest as code; 
crawling from Stack Overflow (e.g., StaQC \cite{yao2018staqc} and CoNaLa \cite{yin2018learning}), which consists of questions and code answers in posts; 
and actual user queries collected from search engines (e.g., CoSQA \cite{huang2021cosqa}), with codes annotated by annotators . 
Each of these data sources has its own limitations: 
in the Github data, code comments are significantly different from actual queries; 
in the Stack Overflow data, the answered code is often not a complete function; 
and in the annotated data, a vast amount of background knowledge is required to understand the code, 
making it difficult to guarantee the scale and quality of the data. 
Therefore, 
we believe it is essential to create a more practical dataset for model training and evaluation. 
One potential solution is to collect online user behavior records, such as clicks and stays, 
which would require a highly performing code search engine with a large user base.
Hence, 
there is a potential to use the existing model to build a preliminary code search engine, 
attract a group of users, and collect user data to create a new dataset. 
  \item \textbf{More fine-grained interaction}. 
Current methods of code search have limitations in their ability to model the interaction between query and code. 
These limitations stem from the fact that existing methods only model the interaction at the representation level \cite{cheng2022csrs, xu2021two} and fail to consider cross-modal interaction at the token level. 
Additionally, these methods use a single level for discrimination, which limits the ability to capture hierarchical information. 
Hence, 
the model architecture used for code search still holds potential for improvement. 
First steps toward that goal have been taken, e.g., by Dong et al. \cite{dong2023retriever}.
  \item \textbf{Improving code search efficiency}. 
Deep learning-based code search methods have demonstrated promising results. 
However, previous approaches have predominantly focused on retrieval accuracy while neglecting the consideration of retrieval efficiency. 
If the code search model is intended for real-world online scenarios, 
enhancing retrieval efficiency becomes a challenge that demands immediate attention. 
Addressing this crucial concern is vital to successfully deploy and utilize the model in practical applications. 
Gu et al. \cite{gu2022accelerating} propose promising first steps into this direction.
  \item \textbf{Context-related code search}. 
The current method assumes that functions in a project are independent \cite{gu2018deep, husain2019codesearchnet, huang2021cosqa}, disregarding their relationships within the project. 
However, 
functions in a project are actually interdependent, with function calls and shared variables. 
To accurately model a function's role, its context must be considered. 
Designing a method to model contextual information and efficiently search for functions with context information online is a valuable research direction. 
  \item \textbf{Personalized code search}. 
Software developers possess unique programming habits, 
leading to varying speeds of understanding and preferences for different forms of code for the same function. 
Consequently, 
code search results should be tailored to individual users based on their programming preferences. 
A potential area for research is the implementation of a personalized code search service. 
  \item \textbf{Better code analysis tools}. 
Different data structures like Abstract Syntax Trees, Control Flow Graphs, Program Dependency Graphs, and Data Flow Graphs can assist in comprehending the semantics of code. 
However, 
the absence of useful open source tools to extract these structures impedes the progress of code intelligence research. 
For instance, 
the extraction of program dependency graphs is currently limited to tools designed specifically for the Java programming language. 
This poses a hindrance to research in the field of code intelligence.
Thus, 
developing better open-source code analysis tools for multiple programming languages would be a beneficial direction to further advance related research. 
  \item \textbf{Exploring new search paradigms.}
Recently, large-scale language models used for code generation have provided functionality akin to code search. 
For instance, 
GitHub's Copilot tool \footnote{https://github.com/features/copilot}, backed by Codex \cite{chen2021evaluating} at its core, 
maps natural language descriptions of desired functionalities to corresponding code snippets that provide those functionalities. 
Similarly, 
OpenAI's ChatGPT tool \footnote{https://openai.com/blog/chatgpt}, powered by GPT-4 \cite{GPT42023OpenAI}, maps natural language descriptions of desired functionalities to code snippets that offer the desired functionalities, 
accompanied by explanatory natural language text to enhance developers' understanding and usability. 
In light of the impact of these tools, 
it is imperative that we reconsider the significance and form of code search tasks. 
Exploring new paradigms may usher in a new era of transformation for code search. 
One promising direction worth exploring is cross-fertilization with other code intelligence tasks, 
such as leveraging code search results to assist code completion or code generation.
\end{enumerate}

\section{Conclusion} 
In this survey, 
we have presented a comprehensive overview of deep learning-based methods for the code search task. 
We start by introducing the code search task and outlining the framework of deep learning-based code search method. 
Next, we detail the methods for extracting representations of query and code, respectively. 
Furthermore, we categorize many loss functions about the model training. 
Finally, we identify several open challenges and promising research directions in this area. 
We hope this survey can help both academic researchers and industry practitioners, and inspire more meaningful work in this field.

% \begin{acks}
% \end{acks}

\bibliographystyle{ACM-Reference-Format}
\bibliography{sample-base}

\appendix

\end{document}